\input epsf
\documentstyle[epsfig]{aipproc}
\newcommand{\D}{\displaystyle}

\begin{document}

\begin{flushright}
UR-1513~~~ ER-40685-909
\end{flushright}

\title{Fiber R and D for the CMS HCAL}

\author{H. S. Budd, A. Bodek, P. de Barbaro, D. Ruggiero, E. Skup}
\address{Department of Physics and Astronomy,
University of Rochester, Rochester, NY 14627}

\maketitle

\begin{center}
To be published in the proceeding of
SCIFI97 Conference, \\November 3-6, 1997, Notre Dame, Indiana.
\end{center}

\begin{abstract} 

    This paper documents the fiber R and D 
for the CMS hadron barrel calorimeter
(HCAL). The R and D includes measurements of  fiber flexibility,
splicing, mirror reflectivity, relative light yield, 
attenuation length, radiation
effects, absolute light yield, and transverse tile uniformity. 
Schematics of the hardware for each measurement are shown.
These studies are done for different diameters and kinds of  multiclad fiber.

\end{abstract} 
\nopagebreak

    The CMS HCAL optical design is similar to the
CDF Plug Upgrade optical  design \cite{cdf_end_plug} \cite{note}.
A wavelength shifting (WLS) fiber, containing K27 waveshifter, embedded 
in the tile collects the scintillation light.  Outside the tile, the
WLS fiber is spliced to a clear fiber. The clear fiber takes the
light to   a connector at the edge of the pan. An optical cable
brings the light to the optical readout box. The readout box
assembles the light from layers to towers and brings the light to
the photodetectors.

    The fibers tested for this paper are all multiclad.   The
diameters of the fibers range from 0.83 mm to 1.0 mm. CMS has chosen
fiber of 0.94 mm diameter.  We test four types of Kuraray fiber, 
non-S (S-25), S-35, S-50,  and S (S-70) \cite{kuraray}. 
The most flexible Kuraray
fiber is S type and the least flexible fiber is non-S type. We test
two batches of WLS Bicron fiber, BCF91A-MC\cite{bicron}.  
Batch 1 is an earlier version 
than Batch 2. We test one batch of Bicron clear fiber, BCF-98 MultiClad,
which was made at the same time at Batch 1 WLS fiber.
The waveshifter for all WLS fiber is K27.

    We use R580-17 phototubes.  This tube is a Hamamatsu 1.5 inch
diameter, 10  stage tube with a green extended photocathode. The
R580-17 photocathode and the photocathode for the CMS  photodetectors, HPDs, 
are the same.

          For most measurements a tile excited by a radioactive 
source generates the light. This insures that the spectrum  
of light is the same
for these tests as the spectrum from the CMS HCAL calorimeter. 
We use both a Cs-137 $\gamma$ source and a Ru-106 $\beta$ 
source. The Cs-137 source is
collimated by a lead cone. The widest diameter of the cone is 7.5 cm.  
A picoammeter reads out the phototube. The data aquisition (DAQ) 
program averages 20 measurements from the picoammeter
and creates a pedestal subtracted data file.
The absolute light yield measurement uses the Ru-106 source. Its 
DAQ consists of a 2249A Lecroy ADC triggered by a coincidence of 
two scintillation counters.

The optical connectors used for these measurements  
were developed by DDK \cite{ddk} and CDF Plug Upgrade Group 
\cite{optical_conn} \cite{cdf_upgrade}. Their part numbers
are MCP-10P-1 (0.83 mm fiber), MCP-10P-2 (0.90 mm fiber),  
MCP-10P-3 (1.00 mm fiber), and MCP-10A (the connector housing). 
  
\section*{Fiber flexibility}

    We have studied Kuraray non-S fiber for flexibility by looking
at the change  in light transmission after the fibers are bent. 
Fibers with a change in transmission greater than 
2\% always have cracks or crazing
in the bent  portions. They have light leaking out from these
cracks. Hence, we test for flexibility by looking for light leaking
out of the bend portions.

\begin{table}
\caption{Flexibility of different fibers. Column 1 gives the 
kind of fiber. Column 2 gives the smallest bend diameter
without the fiber damage. Column 3 gives the largest bend
diameter with fiber damage.}
\begin{tabular}{|l|c|c|} 
\noalign{\vspace{-8pt}} \hline 
Fiber type & Fiber not damaged & Fiber Damaged \\ \hline
Kuraray 1 mm non-S     & 2 1/2 in & 2 in    \\
Kuraray 0.94 mm S-35   & 3/4 in   & 5/8 in  \\
Kuraray 0.94 mm S-50   & 5/8 in   & -    \\ 
Kuraray 0.83 mm S      & 5/8 in   & -    \\ 
Bicron  1.00 mm        & 2 in     &   1 1/2 in \\
     \end{tabular}
\label{flexibility}
\end{table}
  
    We test the flexibility by wrapping the fibers around  dowels
and looking for cracks where light leaks out.
Table~\ref{flexibility} gives the result. The test lasted 1/2 year.
If the fiber develops cracks, the cracks appear 1/2 - 2 days 
after the fibers are wrapped around the dowel. The smallest fiber bend
diameter for the HCAL barrel is 3 inches.  Both the Kuraray S-35
fiber and Bicron fiber are flexible enough for CMS HCAL.

\section*{Fiber Splicing}

\begin{figure}
\begin{center}
\caption{CDF automated UV scanner.} 
\epsfxsize=5.7in
\mbox{\epsffile[20 40 560 178]{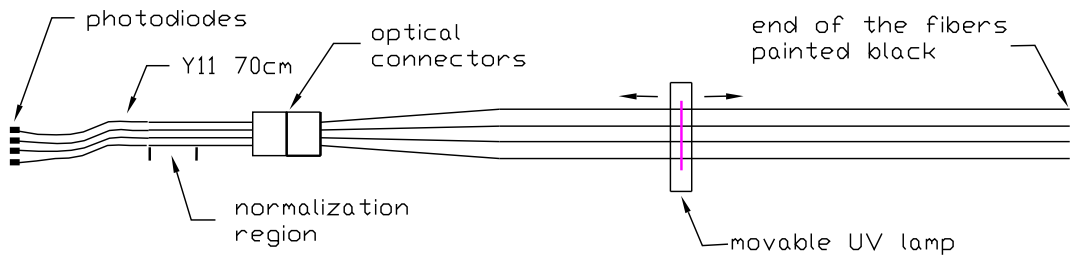}}\\
\label{cdf_uvscanner_eva}
\end{center}
\end{figure}

\begin{figure}
\begin{center}
\caption{ Splice transmission for WLS  fibers. 
(a) compares different kinds of fibers
and (b) compares two different polishing techniques.}
\epsfxsize=2.7in
\mbox{{\epsffile[60 60 517 550]{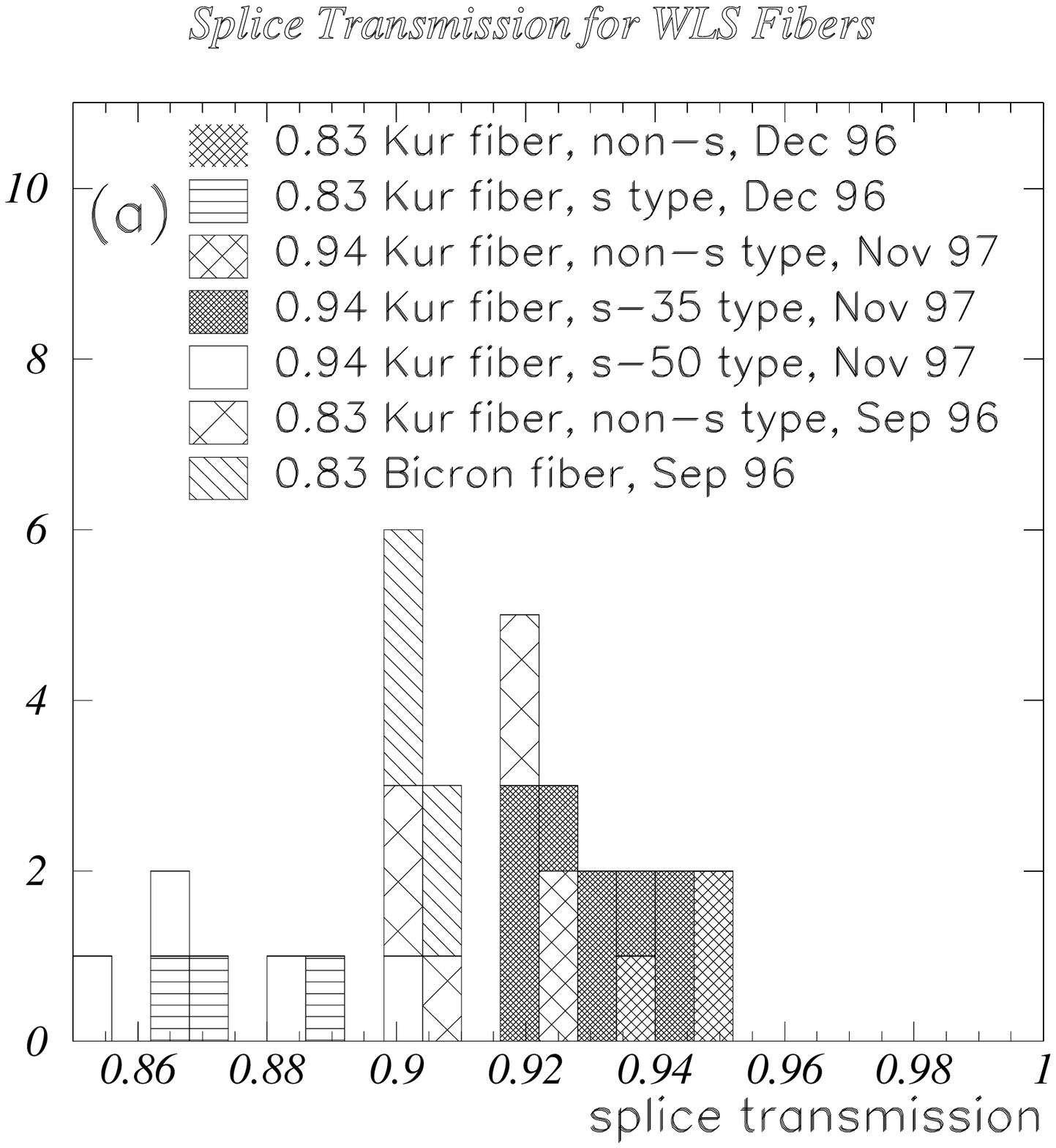}} 
{\epsffile[60 60 517 550]{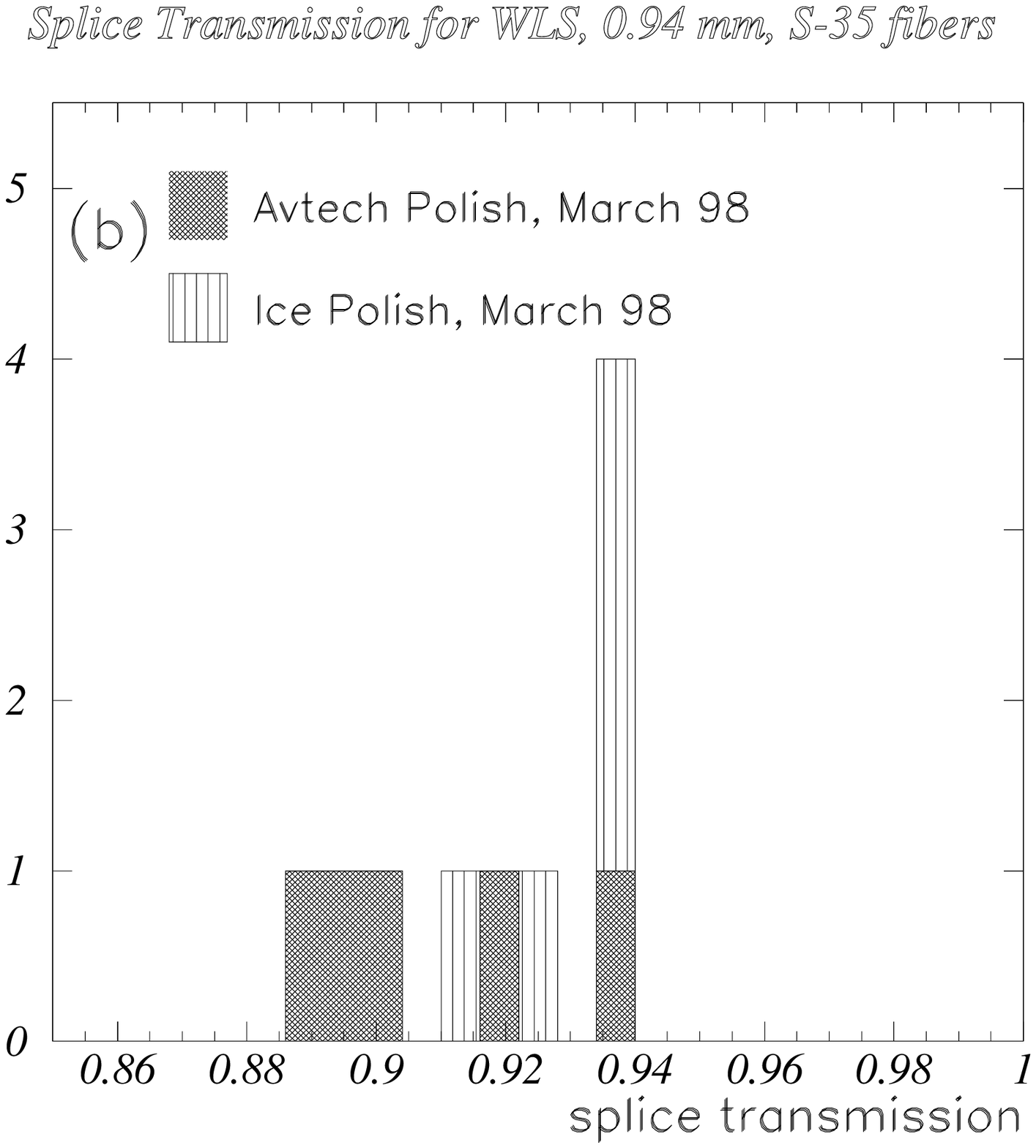}}}\\
\label{splice_green}
\end{center}
\end{figure}

    Fiber splicing is done with the semi-automated splicer developed by
the CDF Plug Upgrade Group~\cite{splicer}. Splice transmission  of WLS
fibers is measured by scanning across a splice using the CDF automated
UV scanner~\cite{splicer}, see  Figure~\ref{cdf_uvscanner_eva}.
Figure~\ref{splice_green}a shows  the results of splicing tests. 
Table \ref{splice} lists the results of the tests. Only
splicing tests done at  the same time should be compared. The Dec~96
splicing test  shows that non-S fiber splices have higher
transmission than S type fiber.  
This result confirms CDF's measurement of the difference 
in splice transmission between non-S and S type fibers~\cite{splicer}. 
The Nov~97 splicing test shows the
splice transmission for non-S and S-35 fiber is the same. Splice
transmission for S-50 is worse than non-S and S-35 fiber. We have
chosen S-35 fiber for the HCAL preproduction prototype because of
its excellent flexibility and high splice transmission. The Sept~96 
splicing test shows that the splice transmission for Kuraray non-S
fiber and  Bicron fiber is the same.  

The fiber ends must be polished well for good splice transmission.
We have compared splice transmission using two different polishing 
techniques. One polishing technique uses the
Avtech polisher \cite{cdf_upgrade}. 
The Avtech polisher is a single fiber polisher
which CDF used for its production. The second technique
is ice polishing. Ice polishing was  pioneered by
Fermilab Charmonium experiment E835 \cite{E835}, and was used by Fermilab
experiment HyperCP E871 \cite{E871}, and D0. The technique involves
freezing fibers in water and polishing the fiber/ice combination.
Many fibers can be polished at once with ice polishing.  
Figure \ref{splice_green}b shows the 
results of splicing Kuraray 0.94 mm, S-35 fibers 
with these two techniques. We conclude that both polishing techniques
give the same transmission. CMS HCAL has chosen to ice polish their fibers.
 
\begin{figure}
\begin{center}
\caption{Setup used to measure transmission of clear
to clear splice. Figure \protect\ref{splice_clear} shows the results.} 
\epsfxsize=5.5in
\mbox{\epsffile[30 80 540 235]{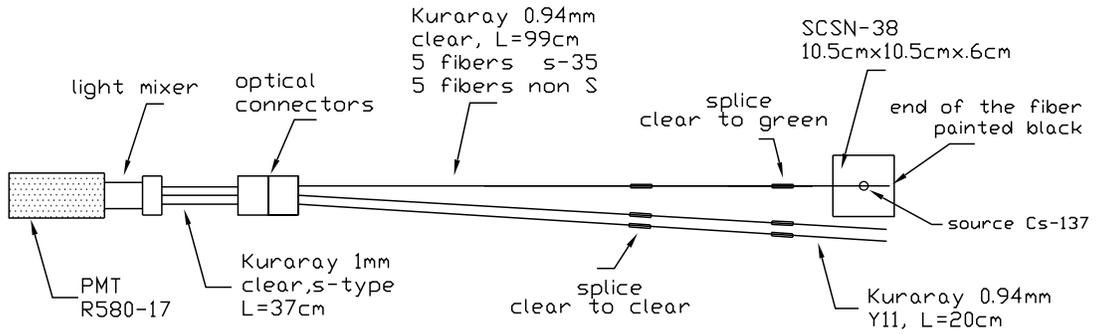}}\\
\label{clear_splice_test_eva}
\end{center}
\end{figure}

\begin{figure}
\begin{center}
\caption{ Splice transmission for clear fibers.}
\epsfxsize=2.7in
\mbox{\epsffile[60 60 517 550]{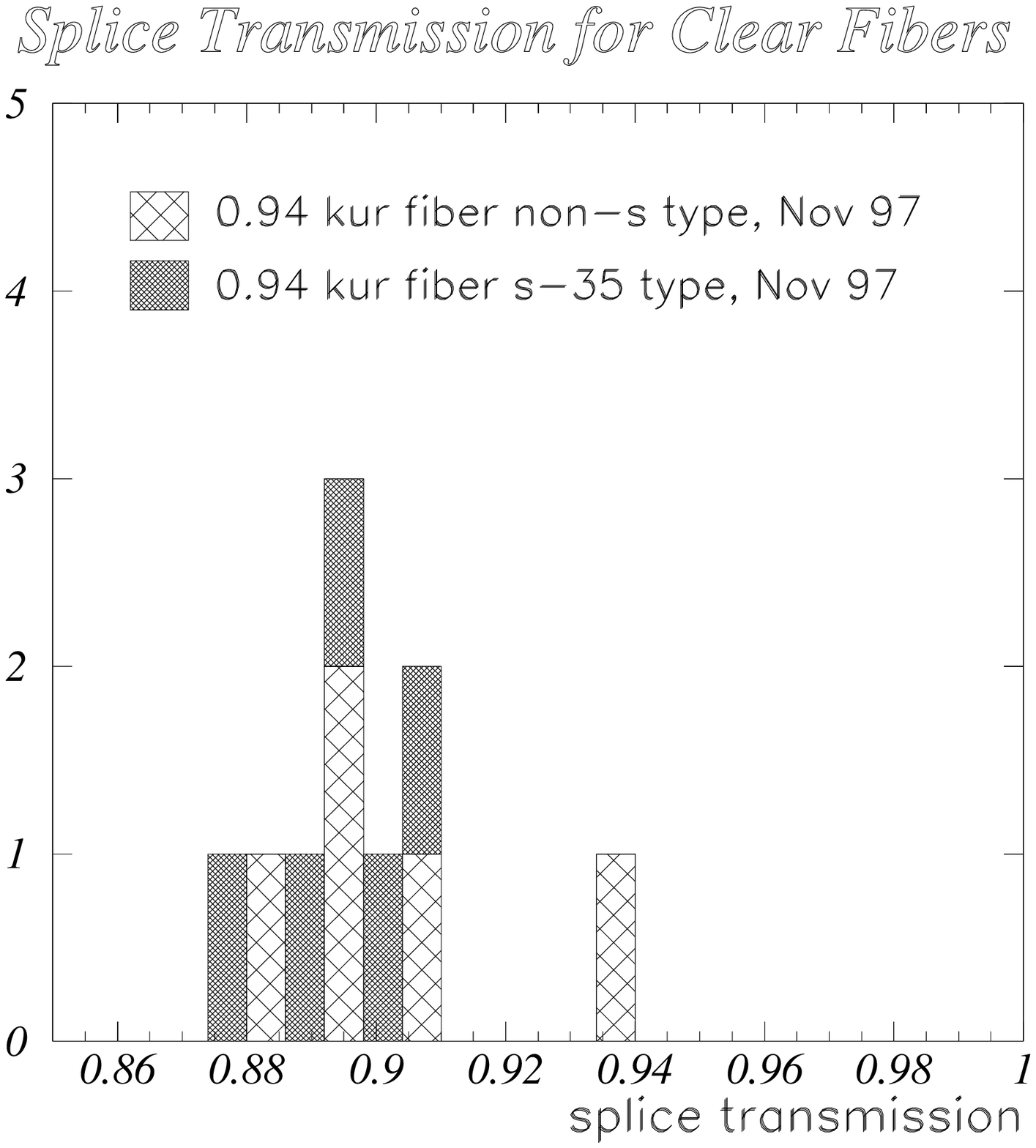}} 
\label{splice_clear}
\end{center}
\end{figure}

    We have tested the splice transmission of  clear fibers. Figure
\ref{clear_splice_test_eva} shows the setup used to
measure the transmission of the splice. 
A connector and fiber assembly with WLS fiber at the nonconnector end  
is called a "pigtail". The WLS fibers in a "pigtail" are 
inserted into a tile and then the tile is excited by a radioactive source 
to readout the light. 
The pigtail for this test consists of  20 cm WLS fibers 
spliced to 99 cm clear fibers in a connector. 
The WLS fiber inserted into a tile injects a constant 
amount of light into the clear fiber. After the  
clear fiber is cut and spliced, the pigtail is remeasured.  
The ratio of the measurement before and after the splice is 
defined as the 
splice transmission. The results are shown   in Figure~\ref{splice_clear}. 
The test shows that  the
splice transmission for clear non-S and clear S-35 fiber is the
same. Both clear and WLS
S-35 splicing tests are done with the same splicing  machine by 
the same operator.  Table \ref{splice} lists all the
fiber splicing results.

\begin{table}
\caption{Splice transmission of different kinds of fibers
and different polishing techniques. Column 1 gives the
kind of fiber. Column 2 gives the fluor. Column 3 gives the
kind of polish. Column 4 gives the date of the test. Column 5
gives the number of fibers tested. Column 6 gives the mean
of the splice distribution. Column 7 gives the RMS of the distribution.
Only tests done on the same date should be compared.}
\begin{tabular}{|l|c|c|c|c|c|c|}
\noalign{\vspace{-8pt}} \hline
Fiber type            & Fluor & Polish &  Date   & Number & Mean & RMS \\ \hline
Kuraray 0.83 mm non-S & K27   & Avtech &  Dec 96 & 3   & 0.948 & 0.006 \\
Kuraray 0.83 mm S     & K27   & Avtech &  Dec 96 & 3   & 0.876 & 0.009 \\
Kuraray 0.94 mm non-S & K27   & Avtech &  Nov 97 & 2   & 0.928 &  --   \\
Kuraray 0.94 mm S-35  & K27   & Avtech &  Nov 97 & 9   & 0.930 & 0.009 \\
Kuraray 0.94 mm S-50  & K27   & Avtech &  Nov 97 & 4   & 0.879 & 0.019 \\
Kuraray 0.83 mm non-S & K27   & Avtech &  Sep 96 & 5   & 0.908 & 0.009 \\
Bicron  0.83 mm       & K27   & Avtech &  Sep 96 & 5   & 0.902 & 0.003 \\
Kuraray 0.94 mm S-35  & K27   & Avtech &  Mar 98 & 5   & 0.908 & 0.018 \\
Kuraray 0.94 mm S-35  & K27   & Ice    &  Mar 98 & 5   & 0.930 & 0.008 \\
Kuraray 0.94 mm non-s & clear & Avtech &  Nov 97 & 5   & 0.904 & 0.018 \\
Kuraray 0.94 mm S-35  & clear & Avtech &  Nov 97 & 5   & 0.893 & 0.011 \\
           \end{tabular}
\label{splice}
\end{table}

\section*{Mirror Reflectivity}

The ends of the fibers are mirrored using
vacuum deposition in Lab 7 at Fermilab. A brief description of the 
mirroring procedure is given in reference \cite{E871}.
We have studied the mirror reflectivity for 2 types of fibers and
both polishing techniques. 

The reflectivity is measured using the automated UV scanner 
(see Figure~\ref{cdf_uvscanner_eva}).
A mirrored fiber is put in a connector and measured with the
UV scanner near the mirror. Next, the mirror is cut off
at 45$^\circ$ to the fiber axis. The end of the fiber is painted
black to prevent any reflection from the end of the fiber.
The fiber is remeasured with the automated UV scanner.
The reflectivity is defined as
(measurement with mirror)/(measurement without mirror) - 1.

\begin{figure}
\begin{center}
\caption{Mirror reflectivity measured with the UV scanner. (a) shows  
the measurement for Kuraray fibers. "Avtech" means the fibers are polished
with the Avtech polisher, and "Ice" means the fibers are ice polished. (b)
compares the reflectivity between Kuraray, 0.83 mm, non-S fiber and 
Bicron 0.83 multiclad fiber. The fibers for (b) used the Avtech polisher.}
\epsfxsize=2.7in
\mbox{{\epsffile[60 60 517 550]{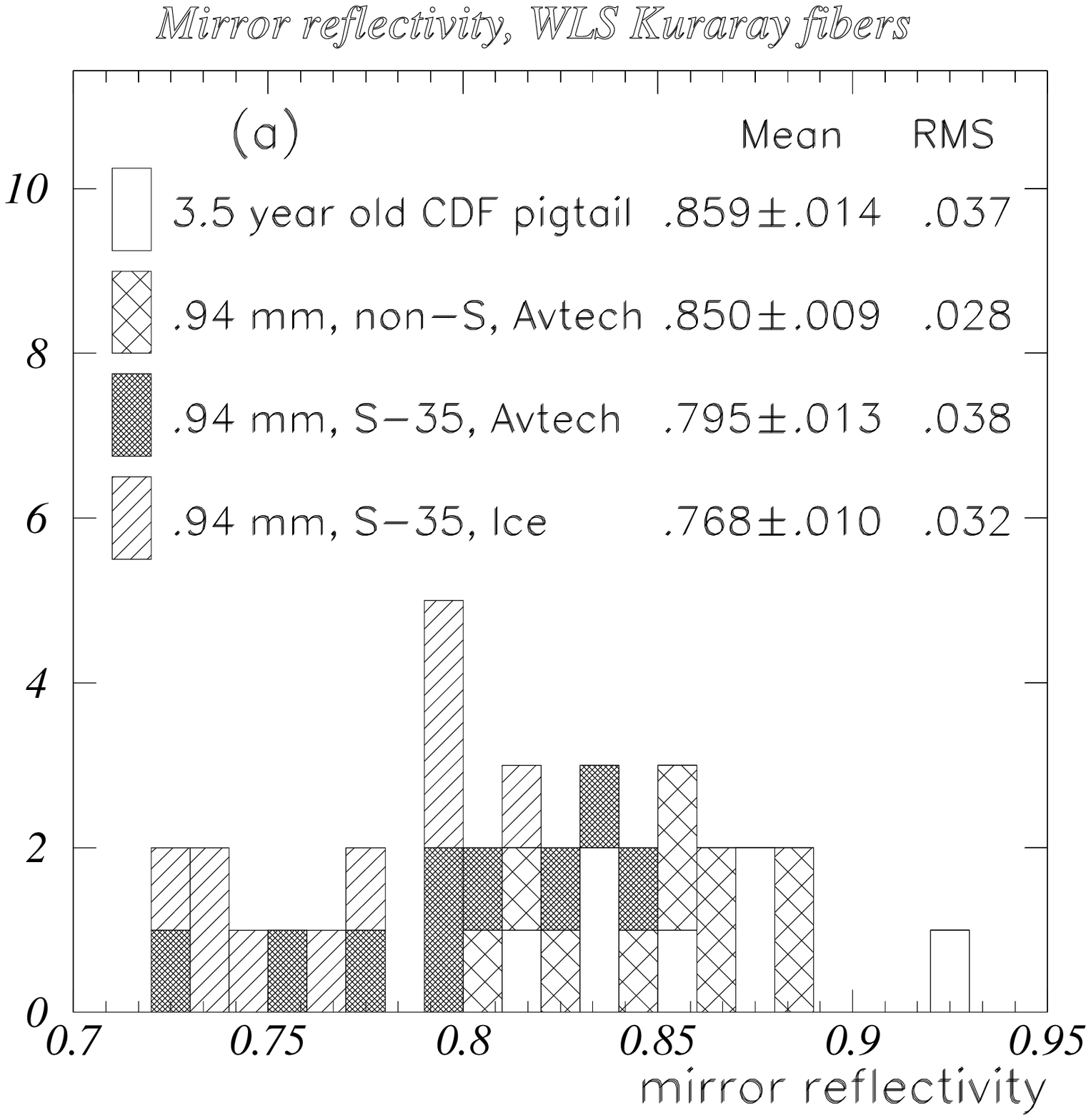}}
{\epsffile[60 60 517 550]{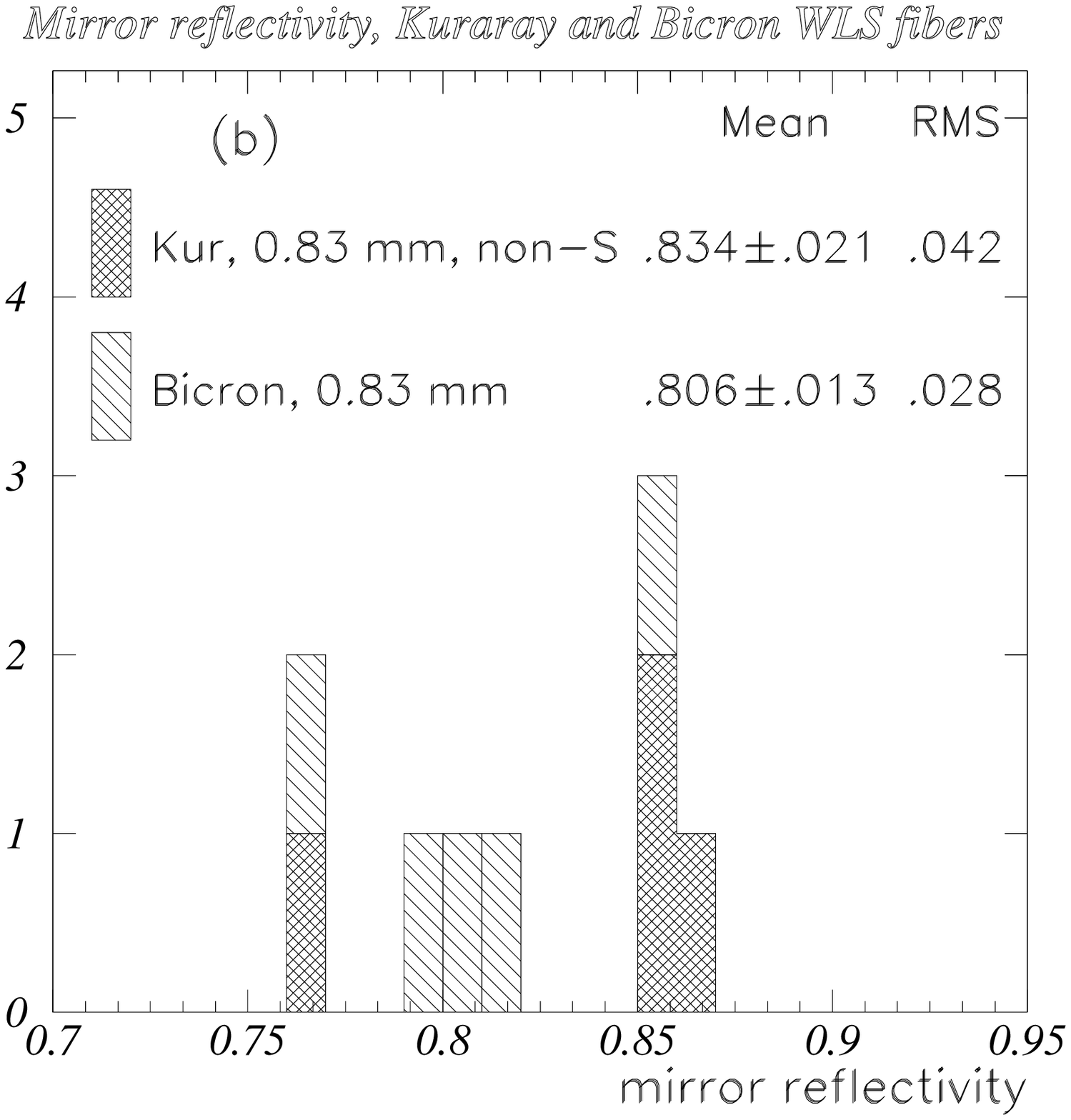}}}\\
\label{mirror}
\end{center}
\end{figure}

Figure \ref{mirror} shows the mirror reflectivity results. We have measured
the reflectivity of a 3 1/2 year old CDF pigtail, which was  
a spare for the CDF End Plug Hadron Calorimeter \cite{cdf_upgrade}. The
pigtail was made of 0.83 mm, non-S fibers polished with the Avtech
polisher. The mirror was dipped in Red Spot UV curable coating 
to protect the mirror.
The measurement shows no degeneration in the mirror
after 3 1/2 years. A measurement
of recently mirrored  non-S type fiber gives the same reflectivity.
The reflectivity of S-35 fibers is roughly 5\% lower.
The reflectivity for ice polished fibers seems to be slightly lower 
than Avtech polished fibers. Figure \ref{mirror}b shows the mirror
reflectivity for Kuraray and Bicron fiber are very similar.

\begin{figure}
\begin{center}
\epsfxsize=5.7in
\caption{Setup to determine the light increase from a mirror in a tile. }
\mbox{\epsffile[30 36 555 250 ]{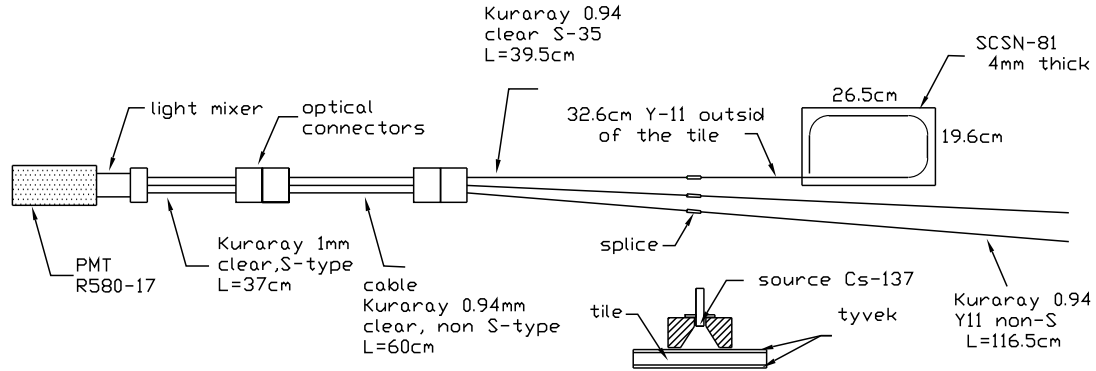}}\\
\label{light_from_mirror_eva}
\end{center}
\end{figure}

We have measured the light increase from the mirror. 
A pigtail with three fibers, shown in Figure~\ref{light_from_mirror_eva}, is made
and scanned with the UV scanner.
The pigtail is also measured with a tile and phototubes as shown 
in Figure~\ref{light_from_mirror_eva}.
Next, the mirror is cut off at 45$^\circ$ to the fiber axis, and  
the end of the fiber is painted black. The pigtail is 
remeasured with both setups. We get the following:
\vspace{-0.2cm}
$$\normalsize \frac{\left( \frac
{\D \text{light from tile with mirror}}
{\D \text{light from tile with no mirror}}\right)}
{ \left( \frac
{\D \text{light from UV scanner with mirror}}
{\D \text{light from UV scanner with no mirror}}\right)}
= \text{0.865 ~~~with RMS = 0.004}$$
This is used to transfer reflectivity measurements using the UV
scanner into reflectivity measurements done using a scintillator tile.
Figure \ref{mirror}a states that light increase from the 
mirror for ice polished S-35 fibers from  the UV scanner is 1.77.
The corresponding increase for a tile is 1.53.

\section*{Relative Light Yield For Different Fibers}

\begin{figure}
\begin{center}
\epsfxsize=5.7in
\caption{Setup for relative light yield test. 
One piece of each hardware is made except for
the 251 cm cables Kuraray 1 mm S type where 2 are made. 
Results are shown in Table \protect\ref{relative_light_yield} and
\protect\ref{ratio_cable_light}.}
\mbox{\epsffile[35 105 555 280]{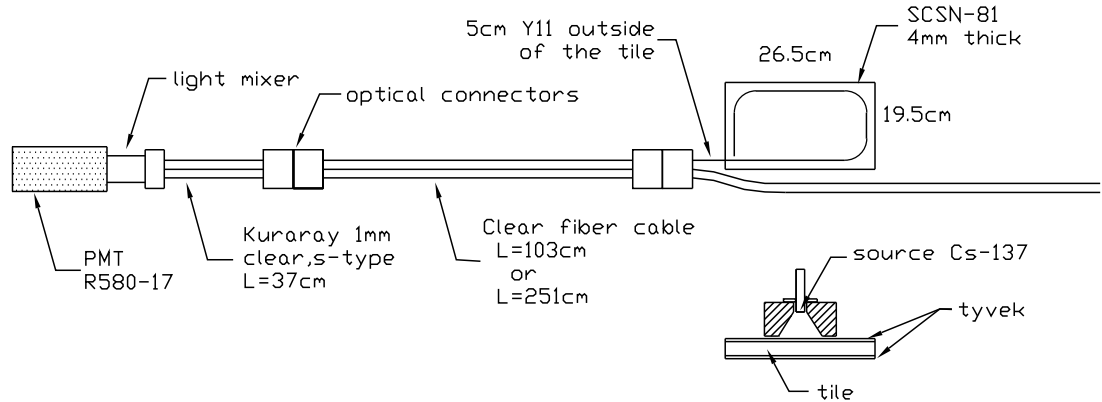}}\\
\label{rel_light_diff_fiber_eva}
\end{center}
\end{figure}

    We measured the light response of different kinds of fibers. 
Figure~\ref{rel_light_diff_fiber_eva} shows the setup used for
the relative light yield test. Fibers are inserted into a tile made of
Kuraray SCSN-81 scintillator. A
clear "cable" connects the pigtail from the connector to the phototube. 
Table~\ref{relative_light_yield}
gives the result for different types of WLS fibers using both 103 cm
and 251 cm Kuraray 1 mm S type clear cables. Column 2 and 3
are separately normalized to the  Kuraray 0.83 mm, 250 ppm, non-S
result (labeled by $\equiv$). For Kuraray fiber, 
the 0.94 mm WLS fiber yields 6\%
more light than a 0.83 mm WLS fiber. 
There is no difference in light between 
a 0.94 mm WLS fiber and a 1.0 mm WLS fiber 
connected to 1 mm clear fiber. Table~\ref{fiber_diameter} 
gives the core diameter and 
fiber diameter. One sample of each fiber is measured. 
The increase in light is smaller than the increase in either the core 
diameter or the fiber diameter.

\begin{table}
\caption{Relative light yield.  The cables are made of 1 mm S 
type Kuraray fiber. Column 2 gives the light using
a 251 cm cable and Column 3 gives the light using a 103 cm cable. 
The entries marked $\equiv$ are defined to be 1.00.}

\begin{tabular}{|l|c|c|}
\noalign{\vspace{-8pt}} \hline 
Type of WLS fiber                  &  251 cm Cable  & 103 cm Cable \\ 
\hline
Bicron 1.00 mm, 200 ppm, batch 1   &        ~~~0.93   &         ~~~0.93  \\
Bicron 1.00 mm, 200 ppm, batch 2   &        ~~~1.00   &         ~~~1.02  \\
Kuraray 0.83 mm, 250 ppm, non-S    & $\equiv$  1.00   & $\equiv$   1.00  \\
Kuraray 0.94 mm, 250 ppm, S-35     &        ~~~1.07   &         ~~~1.06  \\
Kuraray 0.94 mm, 250 ppm, S-50     &        ~~~1.06   &         ~~~1.06  \\
Kuraray 1.00 mm, 200 ppm, non-S    &        ~~~1.06   &         ~~~1.05  \\
Kuraray 1.00 mm, 300 ppm, S        &        ~~~1.08   &         ~~~1.07  \\
     \end{tabular}
\label{relative_light_yield}
\end{table}

\begin{table}
\caption{Core and fiber diameter. All units are mm. 
One fiber is measured for each type. 
$\dagger$ means fiber has only one visible cladding.  
$\ddagger$ means interface between outer cladding
and inner cladding was not distinguishable enough to measure it. 
Kuraray fiber has 2 claddings. Multiclad Bicron 
fiber has one visible cladding, as
the outer cladding is too thin to be visible.}
\begin{tabular}{|l|c|c|c|}
\noalign{\vspace{-8pt}} \hline 
Fiber type                    & Core diameter & Inner cladding diameter & Outside diameter  \\
\hline
Bicron, 0.83 mm, WLS          & 0.789   & $\dagger$  & 0.850               \\
Kuraray, 0.83 mm non-S, WLS   & 0.742   & 0.786      & 0.844               \\
Kuraray, 0.83 mm non-S, clear & 0.737   & $\ddagger$    & 0.838               \\
Kuraray, 0.94 mm non-S, WLS   & 0.841   & 0.903      & 0.959               \\
Kuraray, 0.94 mm non-S, clear & 0.838   & $\ddagger$    & 0.946               \\
Kuraray, 1.00 mm non-S, WLS   & 0.887   & 0.955      & 1.008               \\
Kuraray, 1.00 mm non-S, clear & 0.902   & 0.958      & 1.010               \\
     \end{tabular}
\label{fiber_diameter}
\end{table}

\section*{Attenuation Length of Clear Fibers}

    We have looked at the relative light transmission of different  
clear fibers by using two different cables. Figure
\ref{rel_light_diff_fiber_eva} shows the setup.
Table~\ref{ratio_cable_light} gives the ratio of light for the two
different cables. Column 2 and column 3 give the cables used. The 
length of the cables for column 2 are all 251 cm. Column 4
gives the ratio of the measurement of   column 2 over column 3.
The results are consistent with an  equal attenuation length for 1 mm
S, 0.94 mm S-35,  and 0.94 S-50 Kuraray fiber. 

\begin{table}
\begin{center}
\caption{Comparison of light yields using two different clear cables.}
\begin{tabular}{|l|l|l|c|}
\noalign{\vspace{-8pt}} \hline 
Pigtail            & Cable 1, L=251 cm & Cable 2                  & Cab 1/Cab 2   \\ \hline
Kur 1.00 mm, non-S & Kur 1.00 mm S     & 103 cm Kur 1.00 mm S     & 0.79 \\
Kur 0.94 mm, S-35  & Kur 0.94 mm S-35  & 103 cm Kur 0.94 mm S-35  & 0.80 \\
Kur 0.94 mm, S-50  & Kur 0.94 mm S-50  & 103 cm Kur 0.94 mm S-50  & 0.80 \\
Bicron 1.00 mm     & Bicron 1.00 mm    & 251 cm Kur 1.00 mm S     & 0.89 \\
     \end{tabular}
\label{ratio_cable_light}
\end{center}
\end{table}

\begin{figure}
\begin{center}
\epsfxsize=5.7in
\caption{Setup to measure attenuation length of cables 
of Kuraray, 0.9 mm, S-type clear fibers. Results given in 
Figure \protect\ref{clear_kur_bic}a. 
Pigtails A and B measure different clear fibers.}
\mbox{\epsffile[25 60 555 280]{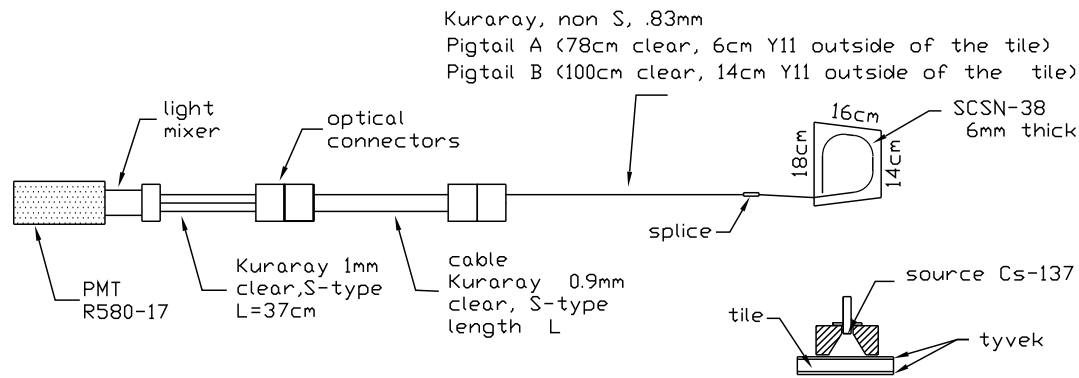}}\\
\label{cable_cdf_att_len_eva}
\end{center}
\end{figure}

\begin{figure}
\begin{center}
\epsfxsize=5.7in
\caption{Apparatus to measure the attenuation 
length of 1 mm clear fiber cables. 
Bicron multiclad and Kuraray S-type fibers are measured. 
The same cables are 
measured with both pigtails. Results given in 
Figure \protect\ref{clear_kur_bic}b.}
\mbox{\epsffile[25 60 555 230]{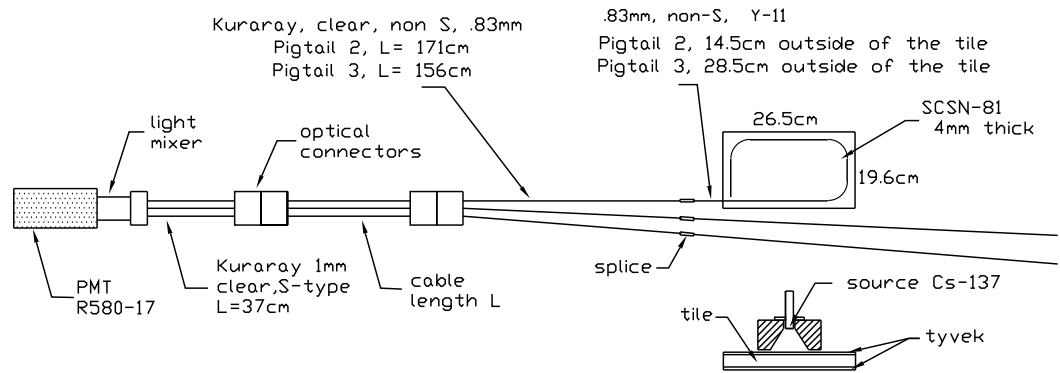}}\\
\label{clear_kur_bic_eva}
\end{center}
\end{figure}

\begin{figure}
\begin{center}
\caption{ Attenuation length of clear fibers.} 
\epsfxsize=2.7in
\mbox{{\epsffile[12 10 510 550]{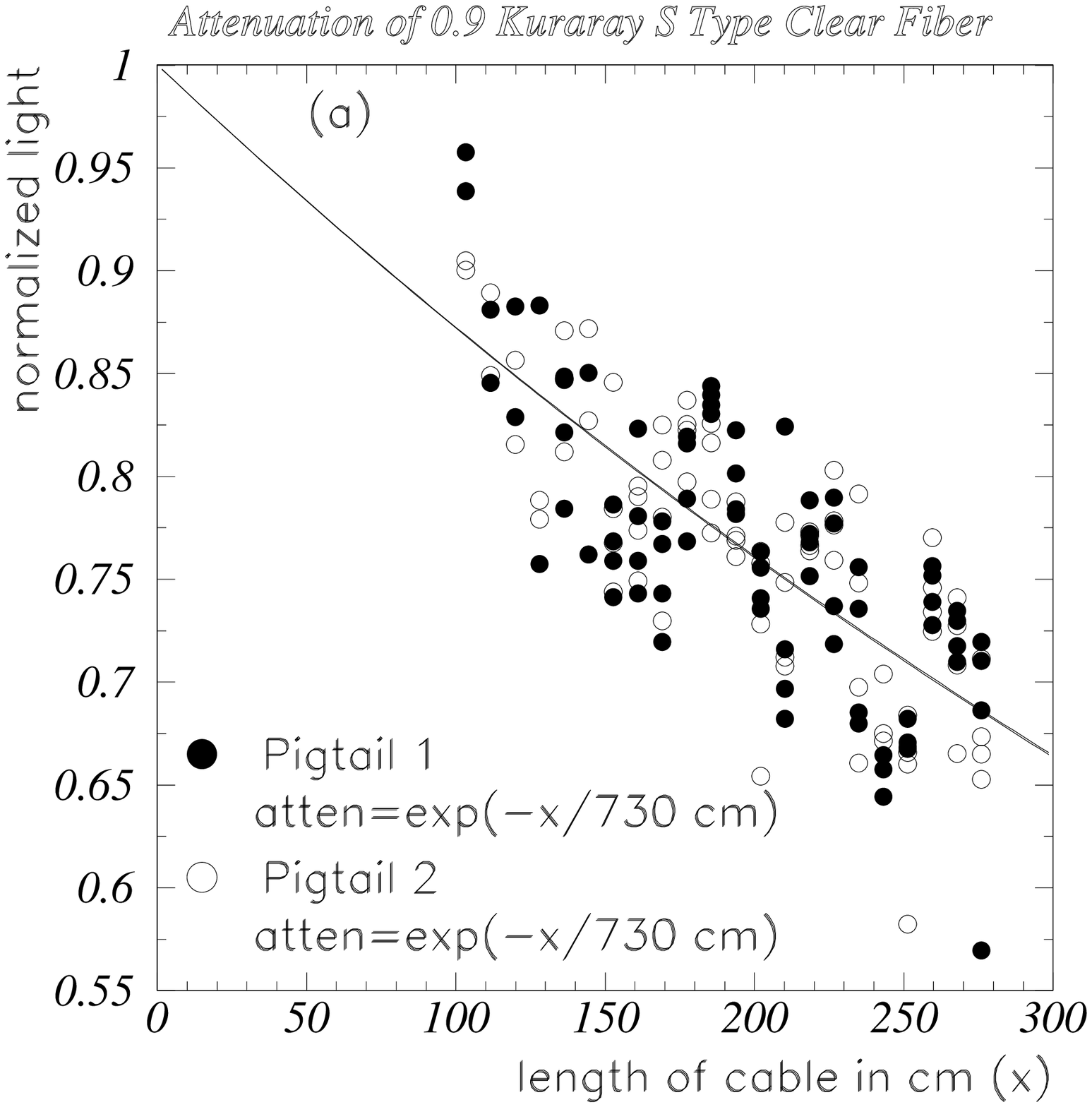}} 
{\epsffile[12 10 510 550]{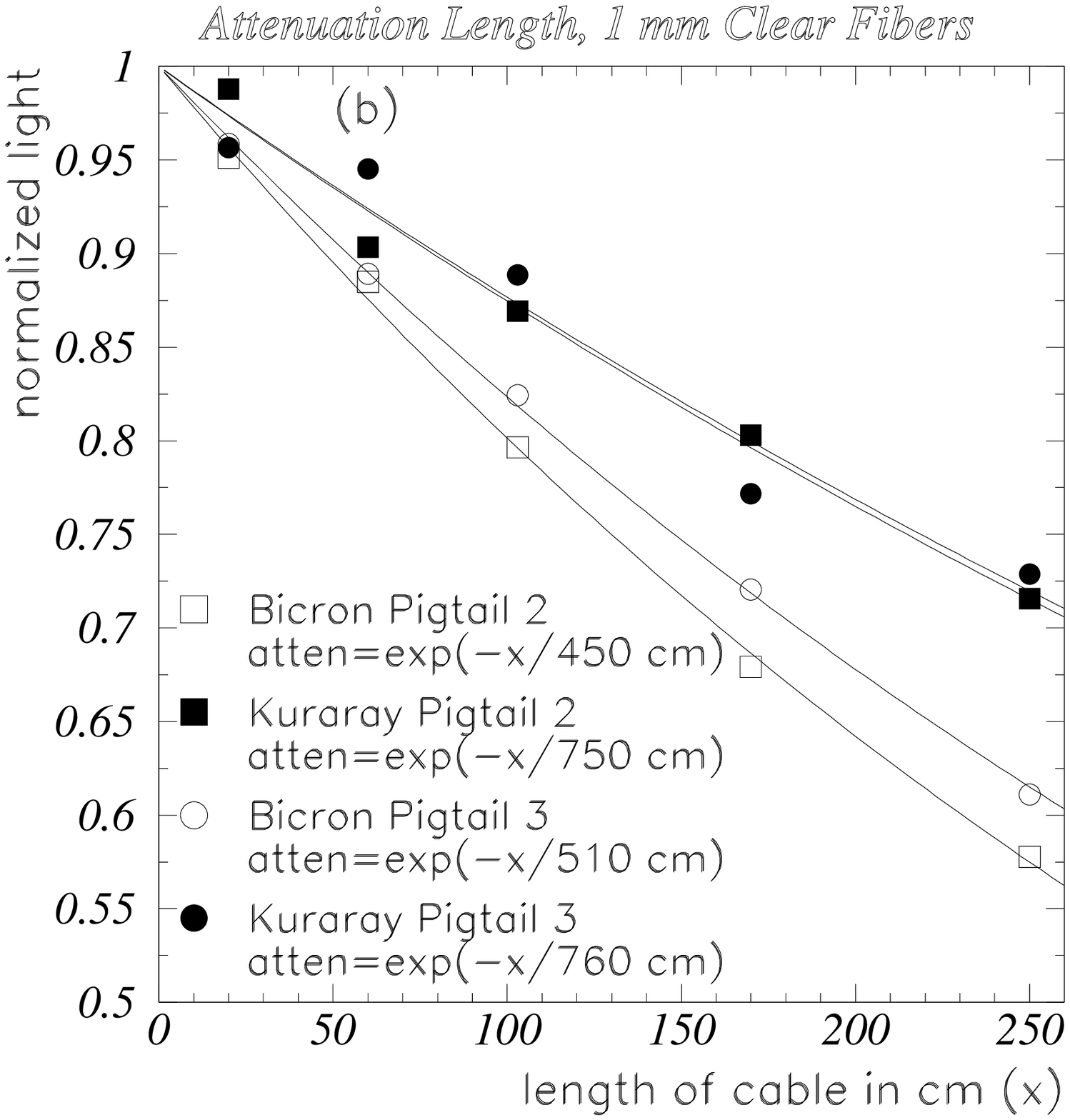}}}\\
\label{clear_kur_bic}
\end{center}
\end{figure}

    We have measured the attenuation length of the clear fiber in  a
cable. Figure \ref{cable_cdf_att_len_eva} gives the apparatus for 
measuring Kuraray 0.9 mm S type fibers, and  Figure
\ref{clear_kur_bic}a gives the result. Each point is a measurement
of one fiber in a cable. Two separate pigtails are  used and they give
the same attenuation length of the  clear fiber.  The combined
results of both pigtails for the  attenuation length is
732 $\pm$ 13 cm. The RMS of the normalized light about the 
exponential curve in Figure \ref{clear_kur_bic}a is
5.6\%.  The test shown in Figure
\ref{clear_kur_bic_eva}   measures the difference in attenuation
between the  Kuraray 1 mm fiber and Bicron 1 mm fiber.  Figure
\ref{clear_kur_bic}b gives the result. Each point is the average of  
the three fibers in the pigtail. The attenuation lengths for Kuraray
fibers given in Figure \ref{clear_kur_bic}a  and Figure
\ref{clear_kur_bic}b agree. 

The data for the clear fiber attenuation measurements are fit to
ae$^{-x/b}$. The data and curve are normalized by setting a = 1. 
A fit to the data used for
Figure \ref{clear_kur_bic}b gives  a$_{kur}$/a$_{bic}$  = 1.02. 
a$_{kur}$/a$_{bic}$ is the   
amount of light accepted by the Kuraray clear fibers divided by the
amount of light accepted by the Bicron clear fibers. 
Theoretically, this should be the same as 
ratio of numerical apertures of the two kinds of fibers. 
Hence, Kuraray fiber and Bicron fiber have the same numerical aperture.

\begin{figure}
\begin{center}
\epsfxsize=5.7in
\caption{Apparatus used to measurement of attenuation length 
of clear fiber spliced to WLS fibers. Results given in 
Figure \protect\ref{fib_att_clear}.}
\mbox{\epsffile[40 80 540 240]{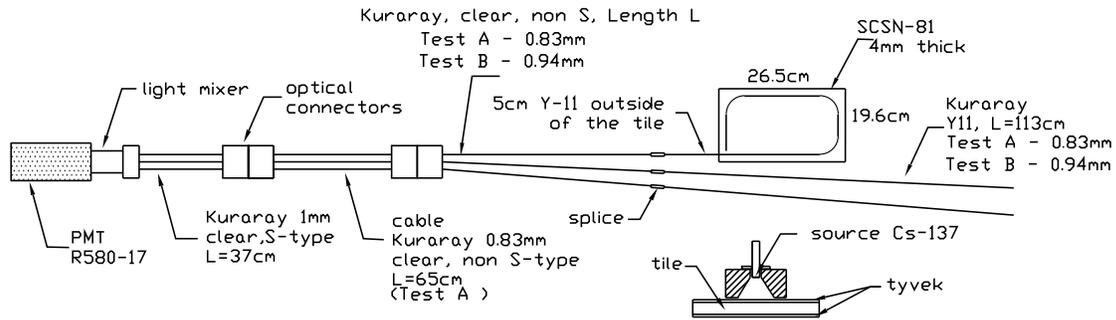}}\\
\label{clear_att_len_eva}
\end{center}
\end{figure}

\begin{figure}
\begin{center}
\caption{Attenuation length of clear fiber spliced to a WLS fiber 
in a pigtail.}
\epsfxsize=2.7in
\mbox{{\epsffile[12 10 510 550]{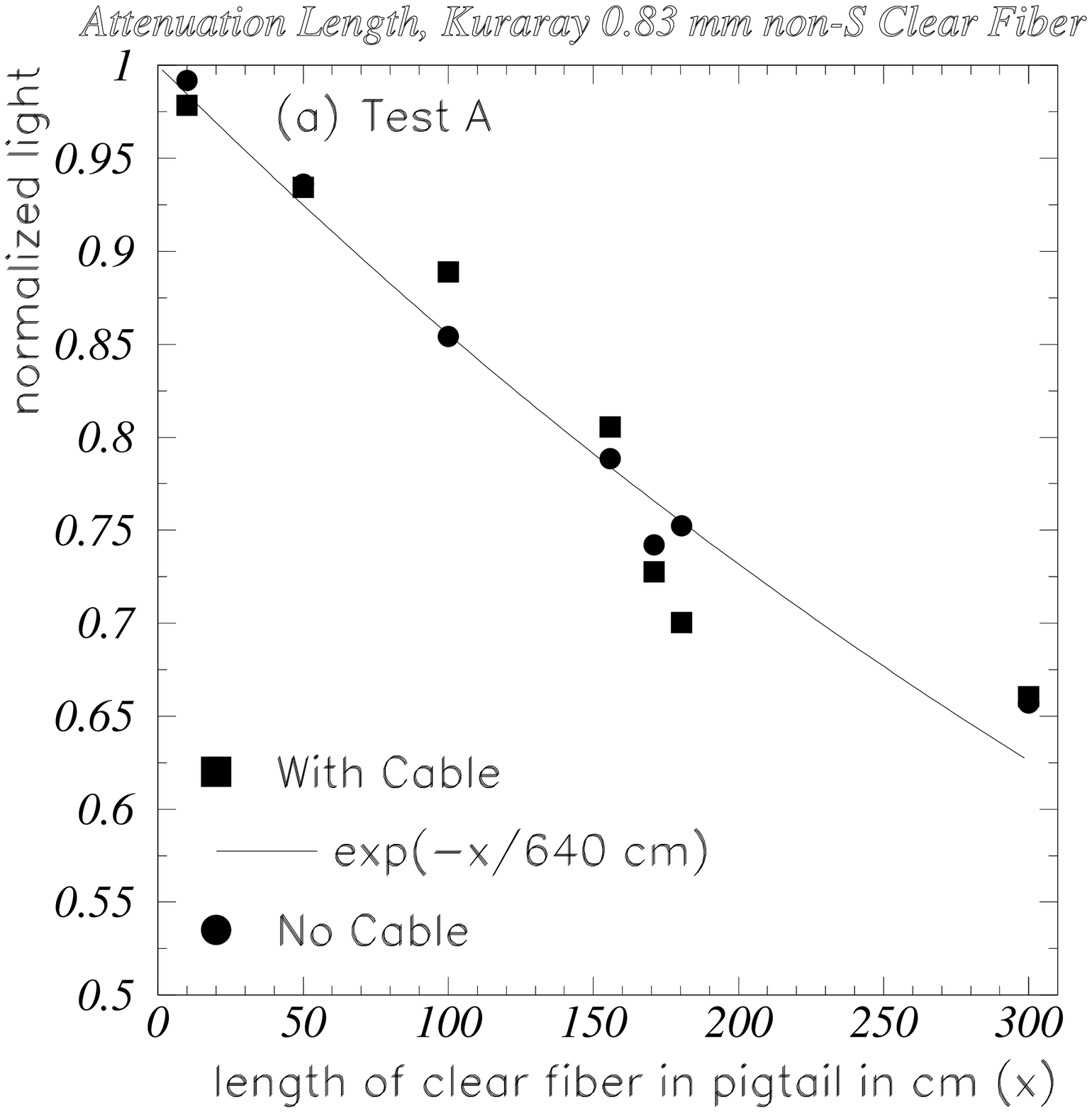}}
{\epsffile[12 10 510 550]{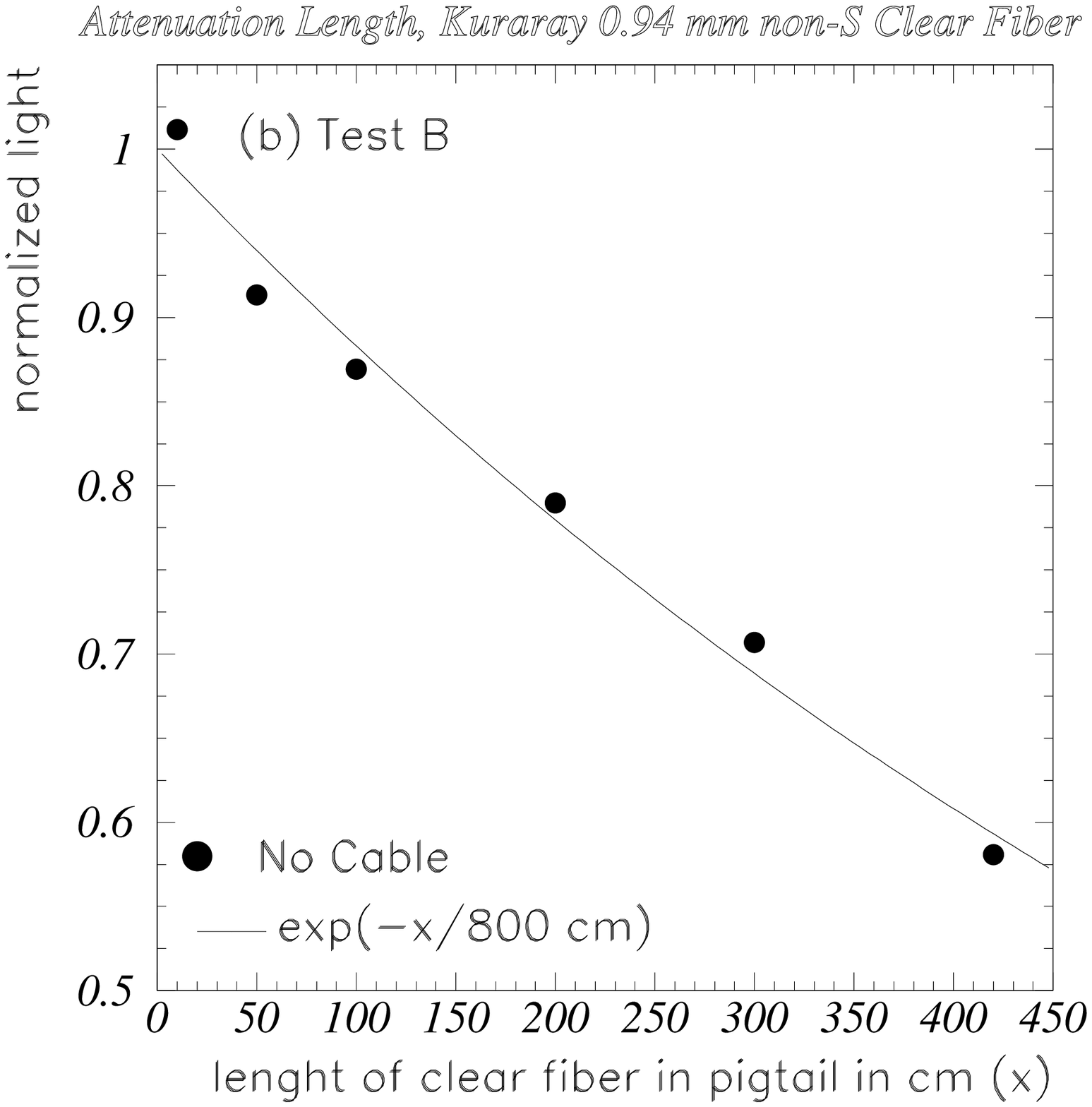}}}\\
\label{fib_att_clear}
\end{center}
\end{figure}

    We have looked at the attenuation length of the clear fiber when
one end is spliced to a WLS fiber and the other end is glued in a
connector. Figure \ref{clear_att_len_eva} shows the apparatus used
to measure the clear attenuation length, and Figure
\ref{fib_att_clear} gives the results. For Test A ( Figure
\ref{fib_att_clear}a), one pigtail with 3 fibers  is  made for each
length of clear fiber.  A single exponential fit gives $\lambda$ = 
6.4 m with the cable and $\lambda$~=~6.55 m without the cable.
Hence, the  attenuation length  of the clear fiber  does not depend
on whether the cable is present.

    For Test B (Figure \ref{fib_att_clear}b), one pigtail was made
with three 4.2 m clear fibers. The  pigtail is measured. The pigtail
connector is cut off and a new connector is put on with the 
clear fiber reduced to 3 m. The pigtail is measured 
and  the process continues until the
pigtail has 0.1 m of clear fiber left.  Test B uses the same splice 
between the clear and
green fibers for all the clear lengths and should give a better
measurement of the attenuation  length. 

The measurement shown in Figure \ref{clear_kur_bic}a gives the best 
measurement of the attenuation length of the clear Kuraray fiber. 
A single exponential with
$\lambda$ = 7.3 m gives an adequate description of the attenuation
length of all Kuraray clear fiber for lengths $<$ 4 m. 
The other measurements of 
the clear Kuraray fiber are consistent with this measurement. 

\section*{Attenuation Length of WLS Fibers}

\begin{figure}
\begin{center}
\caption{Attenuation length of different WLS fibers. The results 
measured with the MINOS setup are shown in (a), and the results
measured with the UV scanner shown in (b). The numbers after 
the symbols in the plots state the number of fibers measured.}
\epsfxsize=2.7in
\mbox{{\epsffile[12 10 510 550]{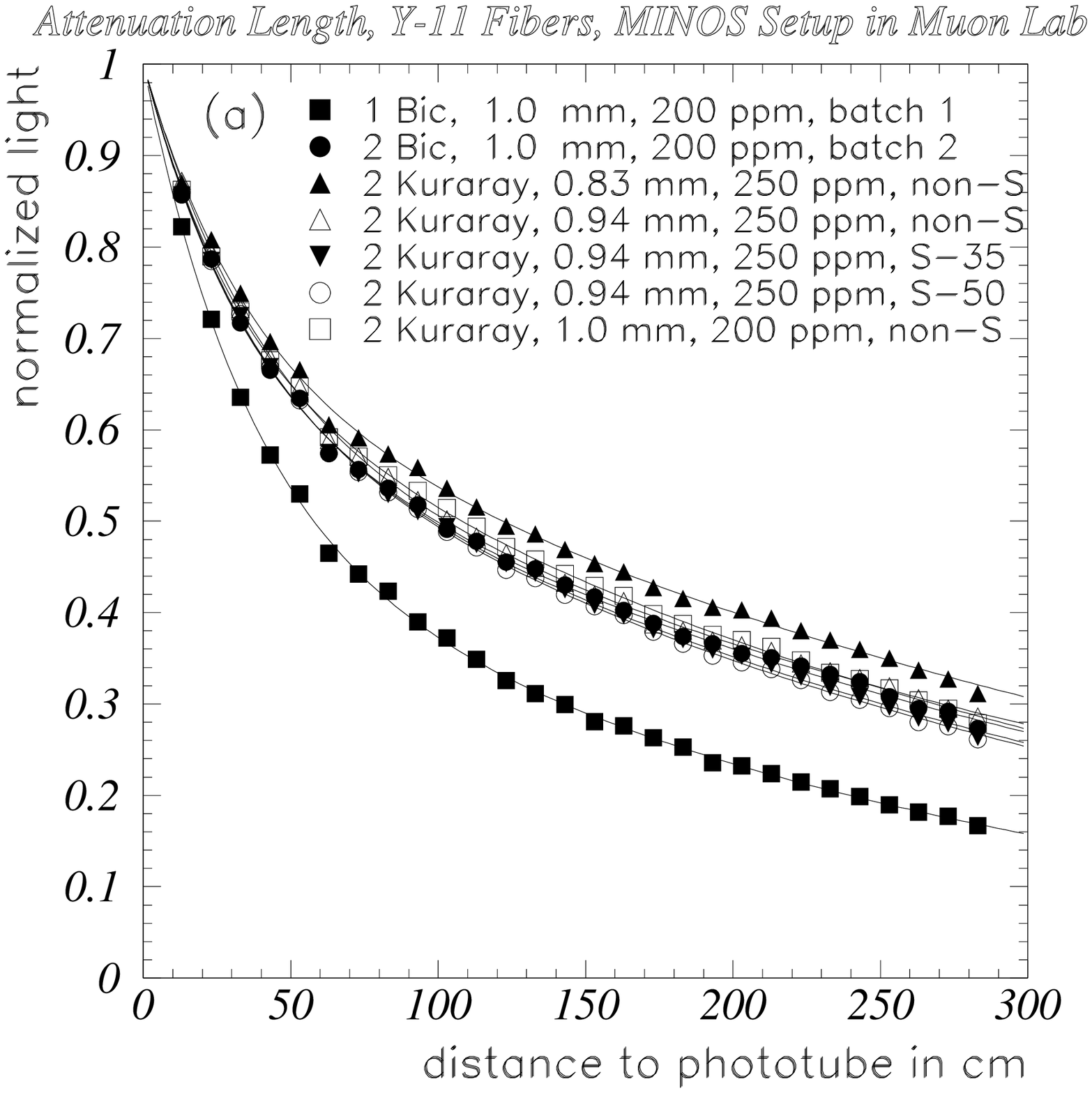}}
{\epsffile[12 10 510 550]{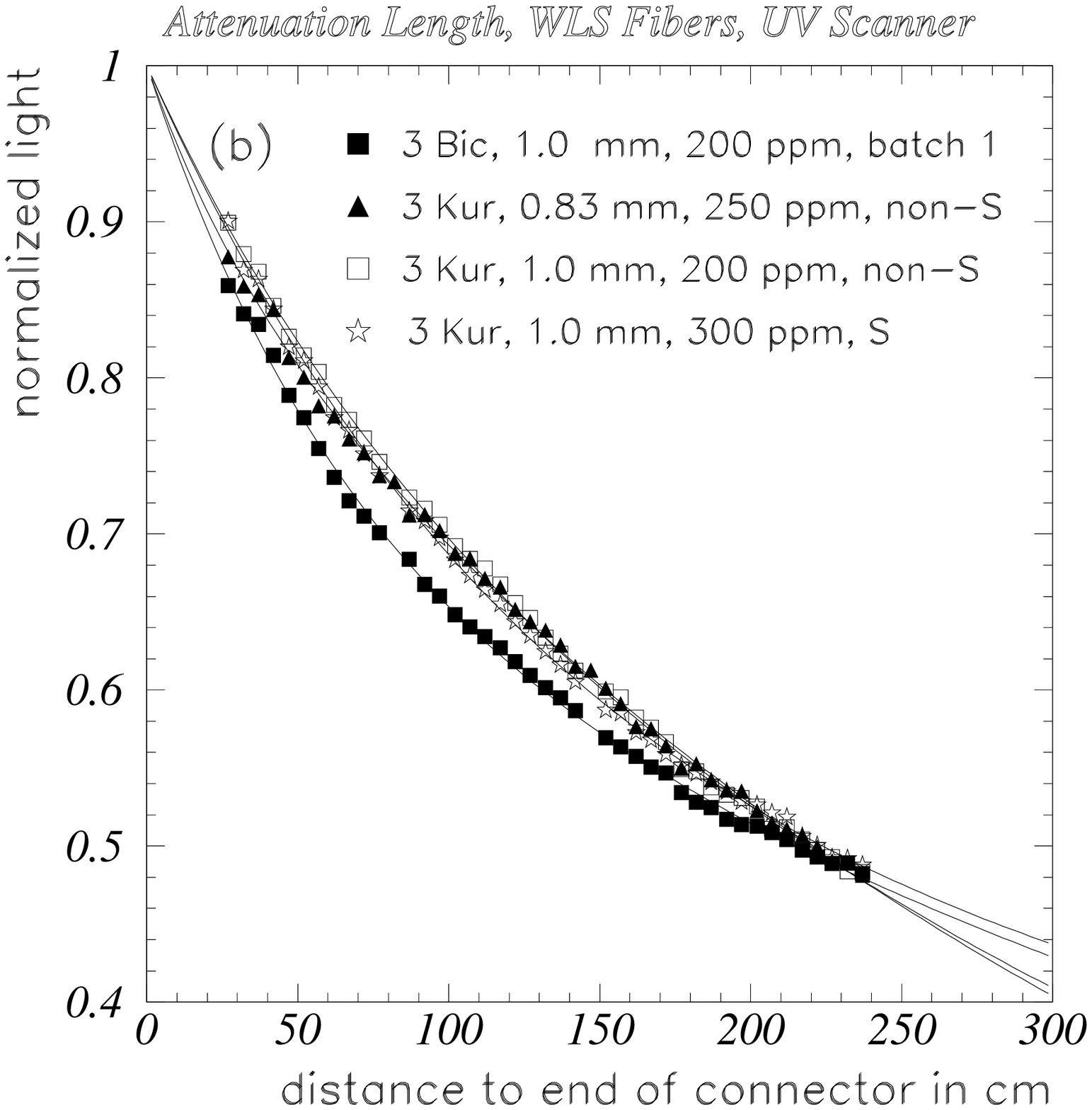}}}\\
\label{fib_att_all}
\end{center}
\end{figure}

\begin{figure}
\begin{center}
\epsfxsize=5.7in
\caption{A schematic of the MINOS setup to measure fiber attenuation.}
\mbox{\epsffile[20 45 560 162]{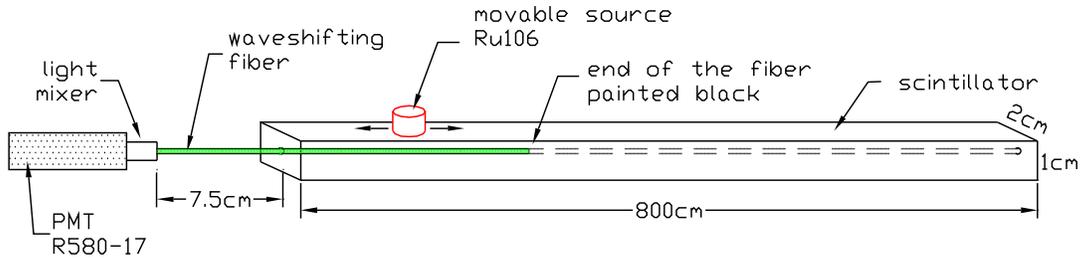}}\\
\label{minos_app_eva}
\end{center}
\end{figure}

    We have measured the attenuation lengths of various WLS fibers.
The attenuation lengths are measured with a setup provided by the
MINOS experiment, see  Figure~\ref{minos_app_eva} . The MINOS setup
provides an easy and quick way of measuring  the attenuation length of
many WLS fibers using scintillator material as a light 
source. One end of the fiber 
is polished, and the other end is cut at 45$^\circ$ and painted black. 
The fiber is inserted into a long hole in the scintillator and the 
polished end is pushed up against a light mixer on the phototube. 
The source is moved across the scintillator and the phototube
current is read out with a picoammeter.  The data are fit to
$ae^{-x/l_1}+be^{-x/l_2}$. Figure~\ref{fib_att_all}a plots the
distance from source to the phototube vs the normalized light.
Table~\ref{K27_atten} gives the numerical results of the fits.

\begin{figure}
\begin{center}
\caption{Comparison of the  attenuation length of Kuraray 0.94 mm, 
250 ppm, non-S measured with 3 m and 4 m pieces.}
\epsfxsize=2.7in
\mbox{\epsffile[ 12 10 510 550]{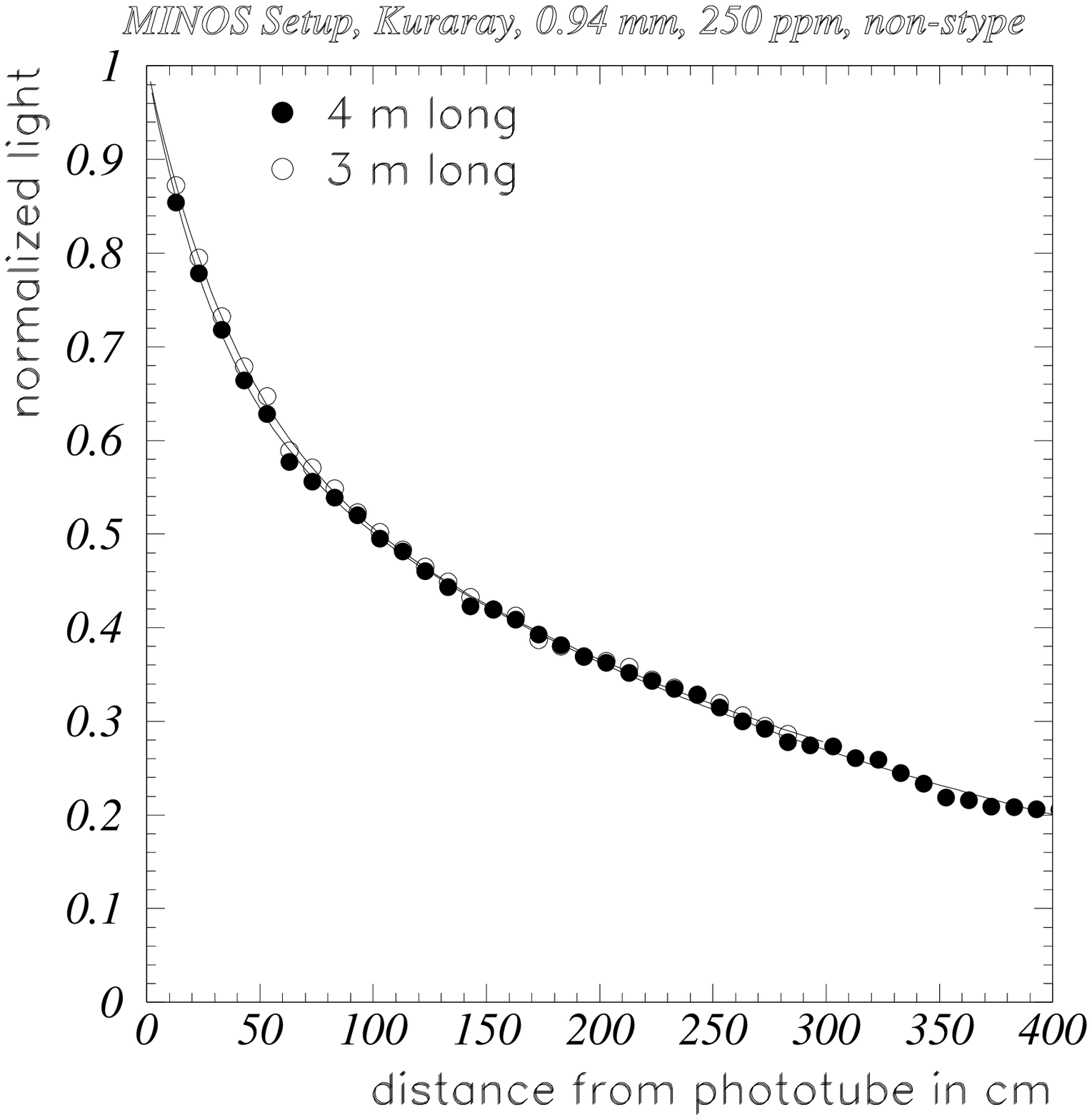}}
\label{fib_att_94_nons_4m}
\end{center}
\end{figure}

    Almost all of our WLS fibers are 3 m long. We have some 
4 m Kuraray 0.94 mm, 250 ppm, non-S fiber. To  determine if the attenuation
lengths using 3 m and 4 m pieces agree,
we measure two 4 m pieces in the MINOS setup. We cut the 2 pieces to
3 m and remeasured them. Figure \ref{fib_att_94_nons_4m} plots
the results. The measurement shows that an attenuation length 
measured with a 3 meter piece can be extrapolated to 4m.  

\begin{table}
\caption{Results of fit to attenuation data for WLS fibers.
The data are fit to $ae^{-x/l_1}+be^{-x/l_2}$. All units for $l_1$ 
and $l_2$ are cm. The column marked Setup gives the apparatus
used to measure the fibers. MINOS uses the MINOS apparatus, 
UV used the CDF UV scanner, and SLIDE is the sliding fiber apparatus.
MINOS RAD are the radiated fibers measured with the MINOS setup.
BEFORE RAD are  the same fibers measured 
before irradiation with the MINOS apparatus with the correction function. 
UV RAD are  the same radiated fibers measured with the UV setup.
These are the same fibers labeled UV in the Setup column.}
\begin{tabular}{|l|c|c|c|c|c|}
\noalign{\vspace{-8pt}} \hline 
 WLS Fiber type                  & Setup     &  $a$  & $l_1$ & $b$ & $l_2$\\ \hline 
 Bicron 1.00 mm, 200 ppm, bat-1  & MINOS      & 0.488 & 35  & 0.512 & 254 \\
 Bicron 1.00 mm, 200 ppm, bat-2  & MINOS      & 0.357 & 33  & 0.643 & 343 \\
 Kuraray 0.83 mm, 250 ppm, non-S & MINOS      & 0.323 & 34  & 0.677 & 379 \\
 Kuraray 0.94 mm, 250 ppm, non-S & MINOS      & 0.372 & 39  & 0.628 & 366 \\
 Kuraray 0.94 mm, 250 ppm, S-35  & MINOS      & 0.345 & 33  & 0.655 & 320 \\
 Kuraray 0.94 mm, 250 ppm, S-50  & MINOS      & 0.348 & 34  & 0.652 & 317 \\
 Kuraray 1.00 mm, 200 ppm, non-S & MINOS      & 0.317 & 31  & 0.683 & 326 \\
 Bicron 1.00 mm, 200 ppm, bat-1  &  UV        & 0.304 & 63  & 0.696 & 611 \\
 Kuraray 0.83 mm, 250 ppm, non-S &  UV        & 0.102 & 33  & 0.898 & 375 \\
 Kuraray 1.00 mm, 200 ppm, non-S &  UV        & 0.188 & 79  & 0.812 & 431 \\
 Kuraray 1.00 mm, 300 ppm, S     &  UV        & 0.524 & 131 & 0.476 & 1407 \\ 
 Kuraray 0.94 mm, 250 ppm, S-35  & SLIDE      & 0.287 & 31  & 0.713 & 366 \\
 Bicron 1.00 mm, 200 ppm, bat-1  & SLIDE      & 0.381 & 31  & 0.619 & 303 \\
 Bicron 1.00 mm, 200 ppm, bat-1  & MINOS RAD  & 0.484 & 30 & 0.516 & 259 \\
 Bicron 1.00 mm, 200 ppm, bat-1  & BEFORE RAD & 0.462 & 39 & 0.538 & 332 \\ 
 Kuraray 1.00 mm, 200 ppm, non-S & MINOS RAD  & 0.343 & 33 & 0.657 & 250 \\
 Kuraray 1.00 mm, 200 ppm, non-S & BEFORE RAD & 0.335 & 42 & 0.665 & 361 \\
 Kuraray 1.00 mm, 300 ppm, S     & MINOS RAD  & 0.362 & 31 & 0.638 & 235 \\
 Kuraray 1.00 mm, 300 ppm, S     & BEFORE RAD & 0.331 & 36 & 0.669 & 323 \\
 Bicron 1.00 mm, 200 ppm, bat-1  & UV RAD     & 0.305 & 58 & 0.695 & 358 \\
 Kuraray 1.00 mm, 200 ppm, non-S & UV RAD     & 0.395 & 148 & 0.605 & 332 \\
 Kuraray 1.00 mm, 300 ppm, S     & UV RAD     & 0.733 & 152 & 0.267 & 1056 \\ \hline
\end{tabular}
\label{K27_atten}
\end{table}

    We compared the above measurement with a measurement using a UV
light and pin diodes to read out the fibers, see
Figure~\ref{cdf_uvscanner_eva}. Figure~\ref{fib_att_all}b plots the
attenuation of the fibers vs the distance to the
connector. The MINOS measurement gives a greater difference between
the Kuraray fiber and  batch 1 Bicron fiber than the UV measurement. 
The greater sensitivity of pin diodes in the UV setup to long
wavelengths light  may be the reason.  Table~\ref{K27_atten} gives
the results of the fits in the rows labeled UV.

\begin{figure}
\begin{center}
\caption{Sliding fiber setup. The results are shown in 
Figure \protect\ref{fib_att_slide}.}
\epsfxsize=5.7in
\mbox{\epsffile[20 45 560 233]{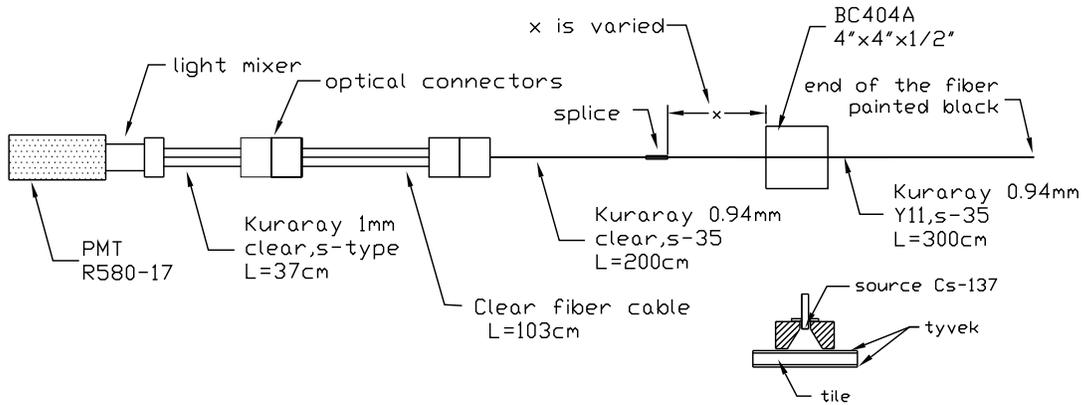}}\\
\label{sliding_fiber_eva}
\end{center}
\end{figure}

\begin{figure}
\begin{center}
\caption{Attenuation measured using sliding fiber setup and compared
with the MINOS measurement. (b) put the x origin for the MINOS measurement
18 cm (Kuraray) and 25 cm (Bicron) into the fiber.}
\epsfxsize=2.7in
\mbox{{\epsffile[12 10 510 550]{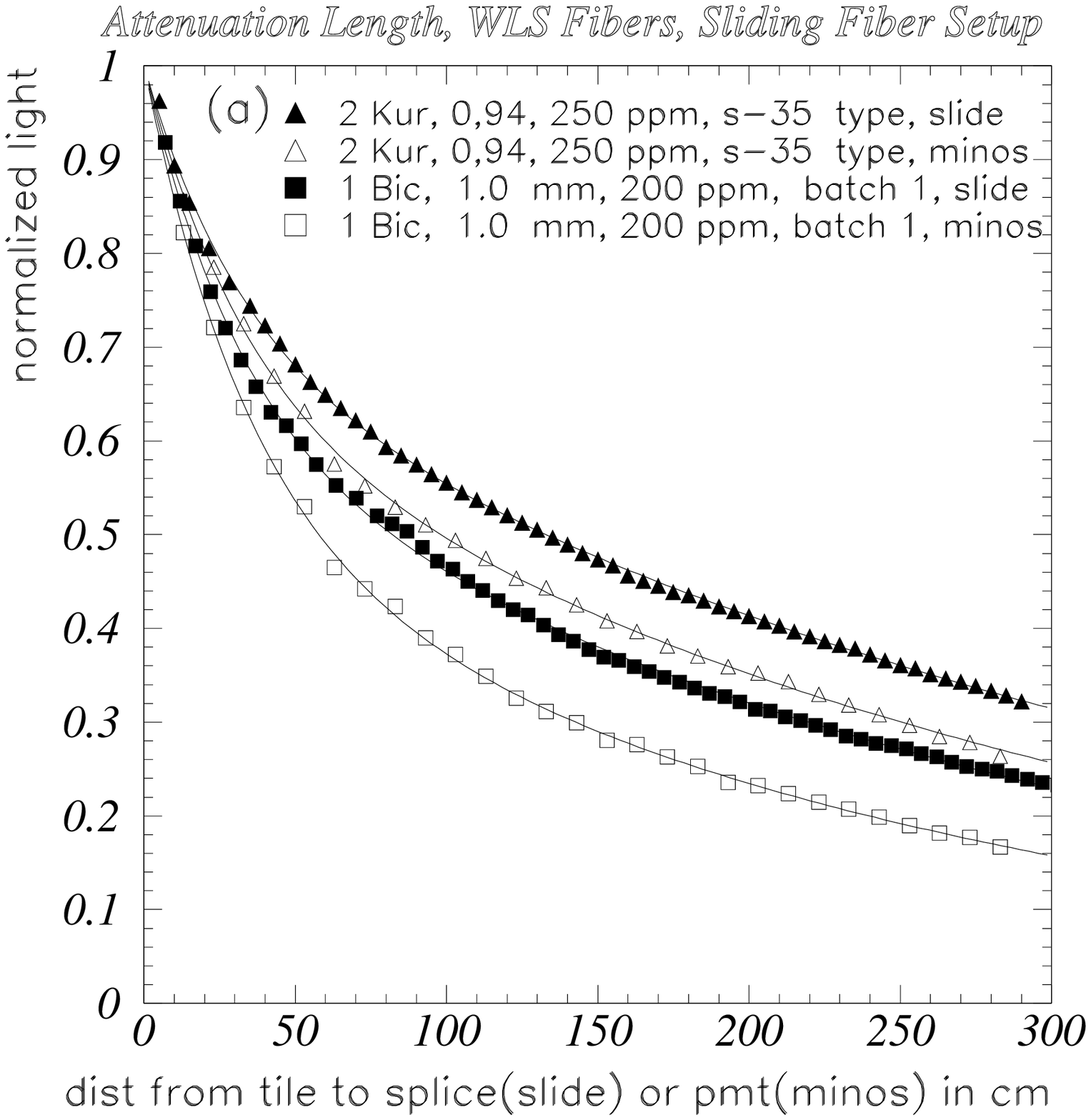}}
{\epsffile[12 10 510 550]{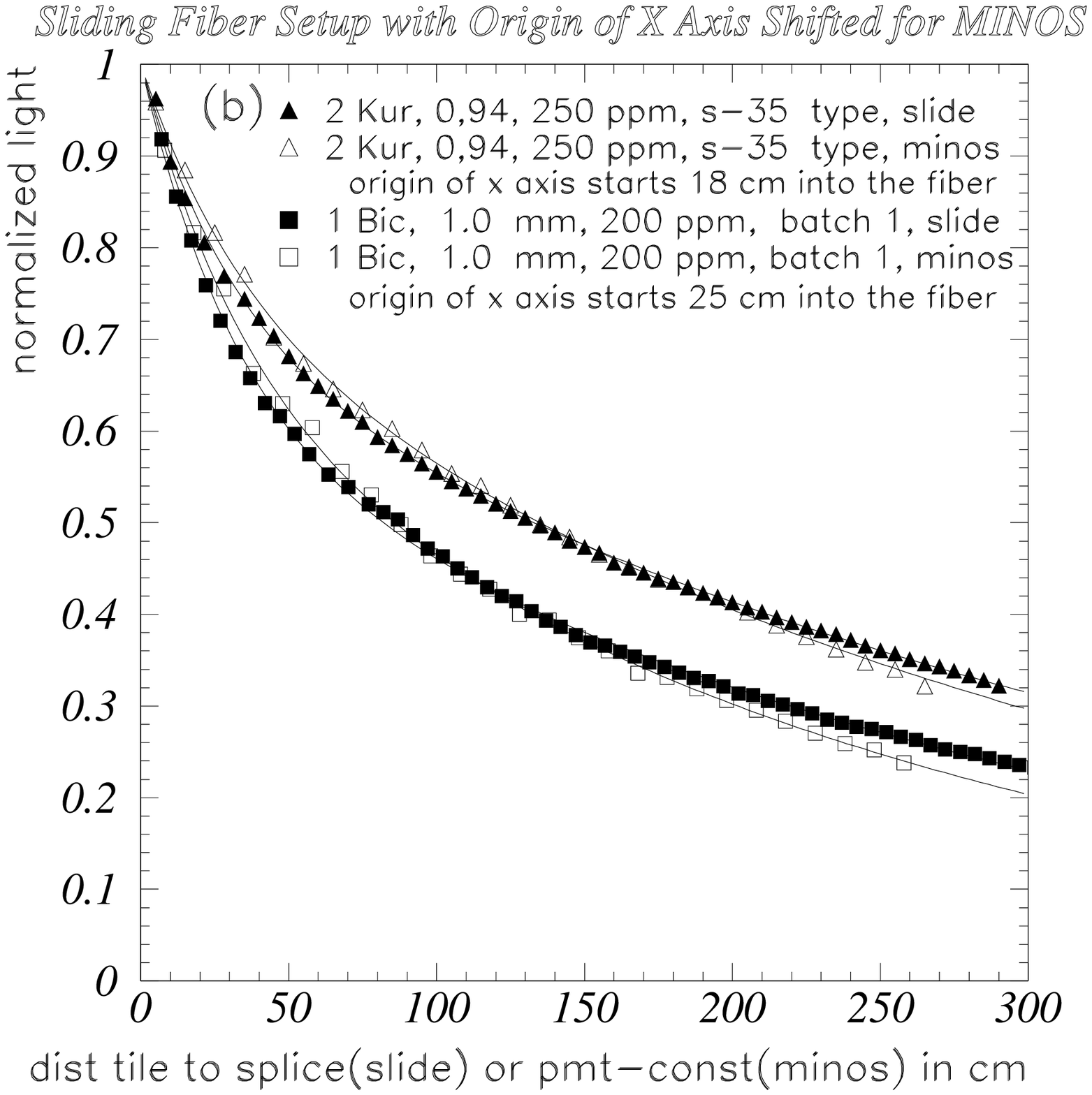}}}\\
\label{fib_att_slide}
\end{center}
\end{figure}

    Fibers can be measured quickly with either the MINOS setup or 
UV scanner. Both tests are useful for comparing fibers. However, the 
setups may not   give the correct attenuation length of the WLS  
fibers which are relevant for CMS HCAL design. 
Figure~\ref{sliding_fiber_eva} shows the setup designed to give a
more accurate attenuation length. Figure~\ref{fib_att_slide}a shows
the attenuation length of two kinds of fibers measured with the
sliding fiber setup. We have shown the measurements of the same 
fibers with the MINOS measurement. Table~\ref{K27_atten} gives the
results for the sliding fiber test for those entries marked SLIDE in
column 2. The results of the sliding fiber setup  for 0.94 mm
Kuraray are used in the design of CMS HCAL. 

    Figure ~\ref{fib_att_slide}a shows the difference in attenuation
length of a fiber measured in two different ways, but with 
the same photodetector and light injection. The difference seen is
due to the extra clear fiber cable and the clear to WLS splice of the sliding 
fiber setup. Figure~\ref{fib_att_slide}b compares the sliding tile
measurement with the MINOS measurement with x=0 set  
18 cm  (Kuraray) and 25 cm (Bicron) away from the phototube
into the scintillator. The
comparison shows  18 cm (Kuraray) of WLS fiber
is acting like the clear fiber and cable.

\section*{Radiation Study of Fibers}

    Some of the WLS fibers were irradiated with 127 krad 
at an electron source at Florida State University.

We encountered a problem in measuring the irradiated fibers. The fibers
were first measured with the MINOS setup in Lab 5 at Fermilab. While 
the fibers were being  irradiated, the MINOS setup was moved to the Muon 
Lab. We measured the irradiated fibers with the MINOS setup in Muon Lab.

\begin{figure}
\begin{center}
\caption{ Comparison of attenuation length of WLS fibers  
before and after irradiation with 127 Krad. The fiber diameters
are 1.0 mm. The 2 setups to
measure the fibers are (a) MINOS setup and (b) CDF UV Scanner.}
\epsfxsize=2.7in
\mbox{{\epsffile[12 10 510 550]{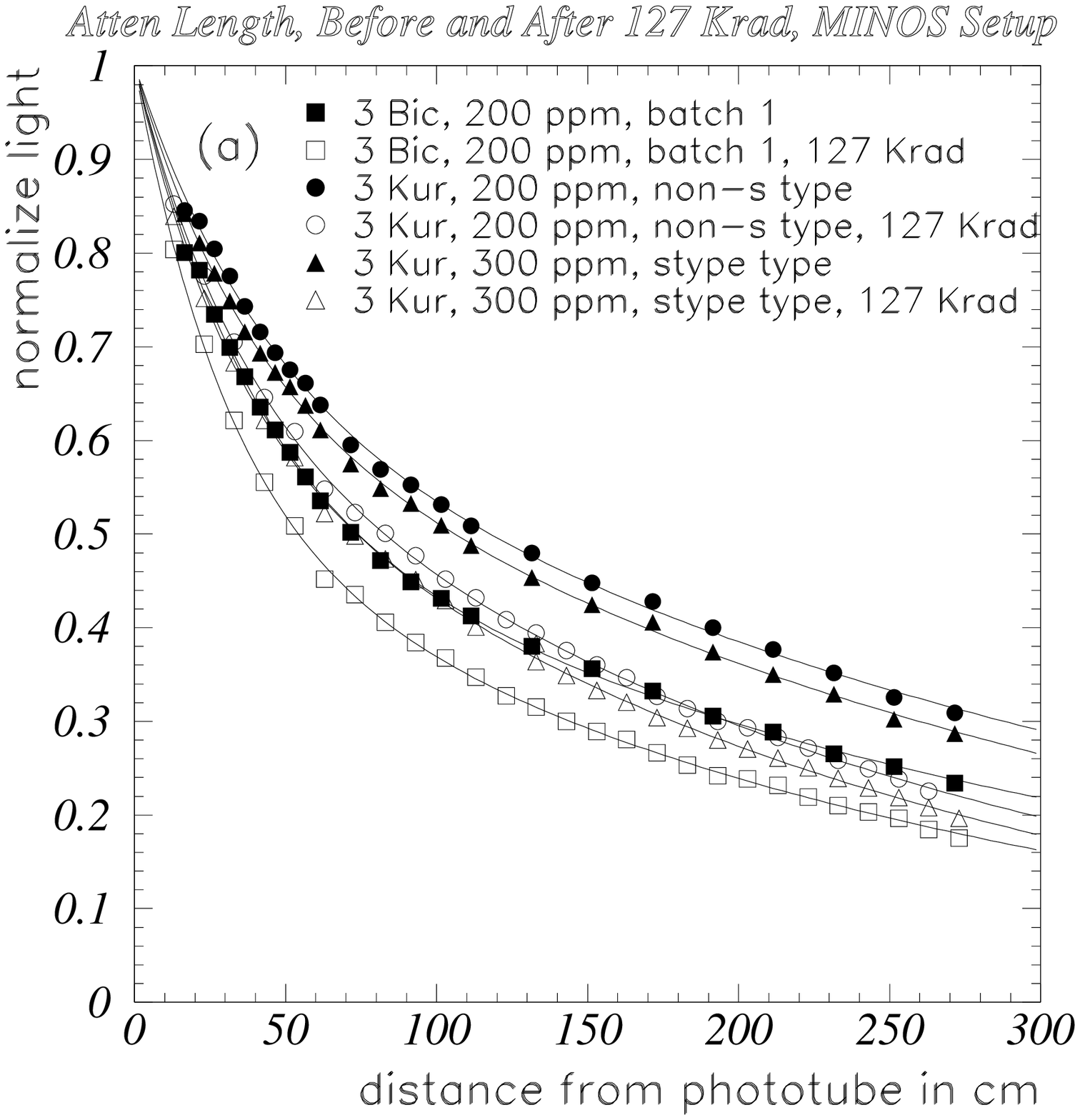}} 
{\epsffile[12 10 510 550]{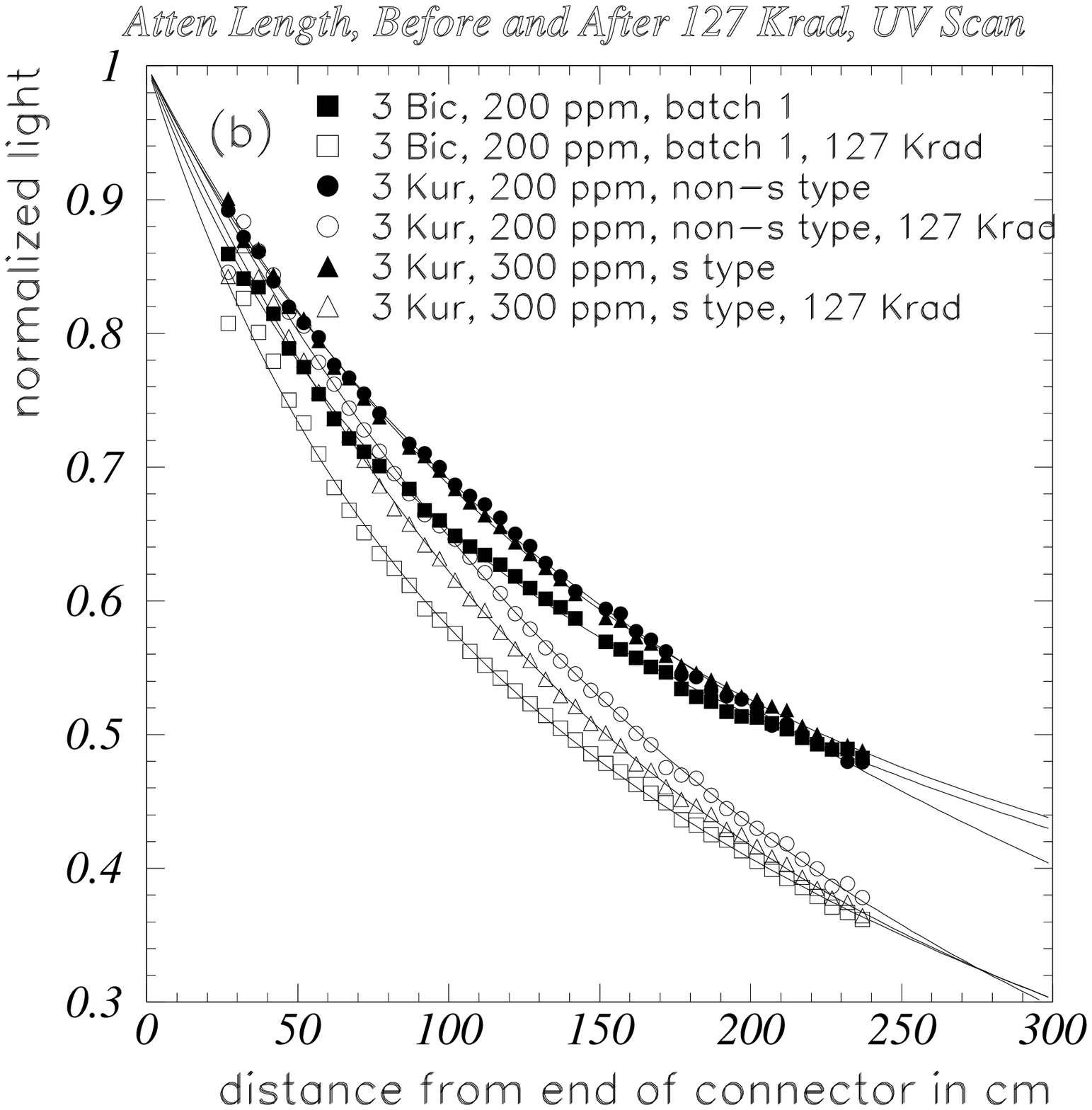}}}\\
\label{fib_att_rad}
\end{center}
\end{figure}

    We measured the same three Kuraray 0.83 mm fibers in both Lab 5 and 
the Muon Lab. The normalized light at 100 cm was .57 in Lab 5 and .52 in
the Muon Lab.  We have no explanation for the difference, since the
setups are the same. The normalized light difference between
different  fiber types is not affected by this problem, but the
absolute normalized light is affected by this problem. The  
measurement of the 0.83 mm fiber in both Lab 5 and Muon Lab is used to get a
correction function for the Lab 5 measurement. The normalized light yield
of the fibers before irradiation is multiplied  by the
correction function. Figure \ref{fib_att_rad} compares 
the normalized light of
the  fibers before and after radiation measured with both the MINOS
setup and the CDF UV scanner. The results of the fits of the fibers 
before and after radiation are given in Table~\ref{K27_atten}.

\section*{Light vs Tile Size and Absolute Light}

\begin{figure}
\begin{center}
\caption{Setup used to measure relative light vs tile size for 
Tower 10. The results are given in Figure \protect\ref{l_over_a}.}
\epsfxsize=5.7in
\mbox{\epsffile[30 70 545 235]{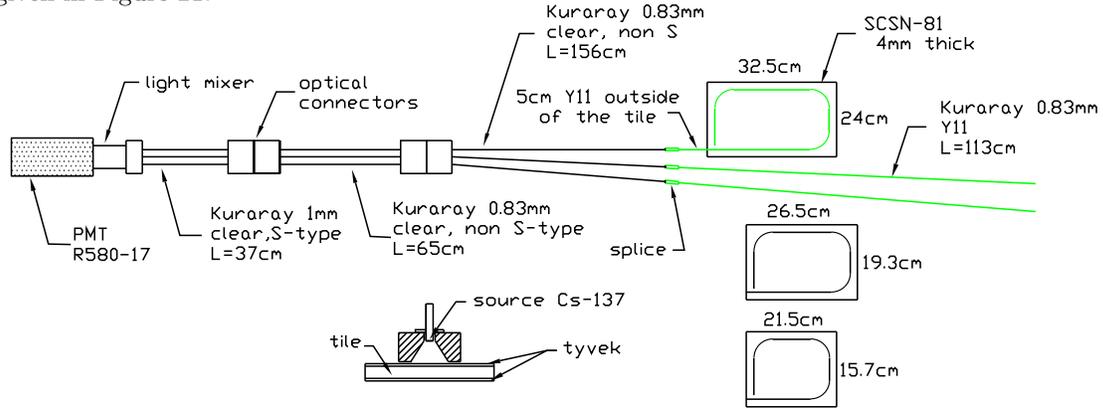}}\\
\label{light_vs_tile_size_lay_10_eva} 
\end{center}
\end{figure}

\begin{figure}
\begin{center}
\caption{Setup used to measure relative light vs tile size for 
Tower 16. The results are given in Figure \protect\ref{l_over_a}.}
\epsfxsize=5.7in
\mbox{\epsffile[35 50 540 250]{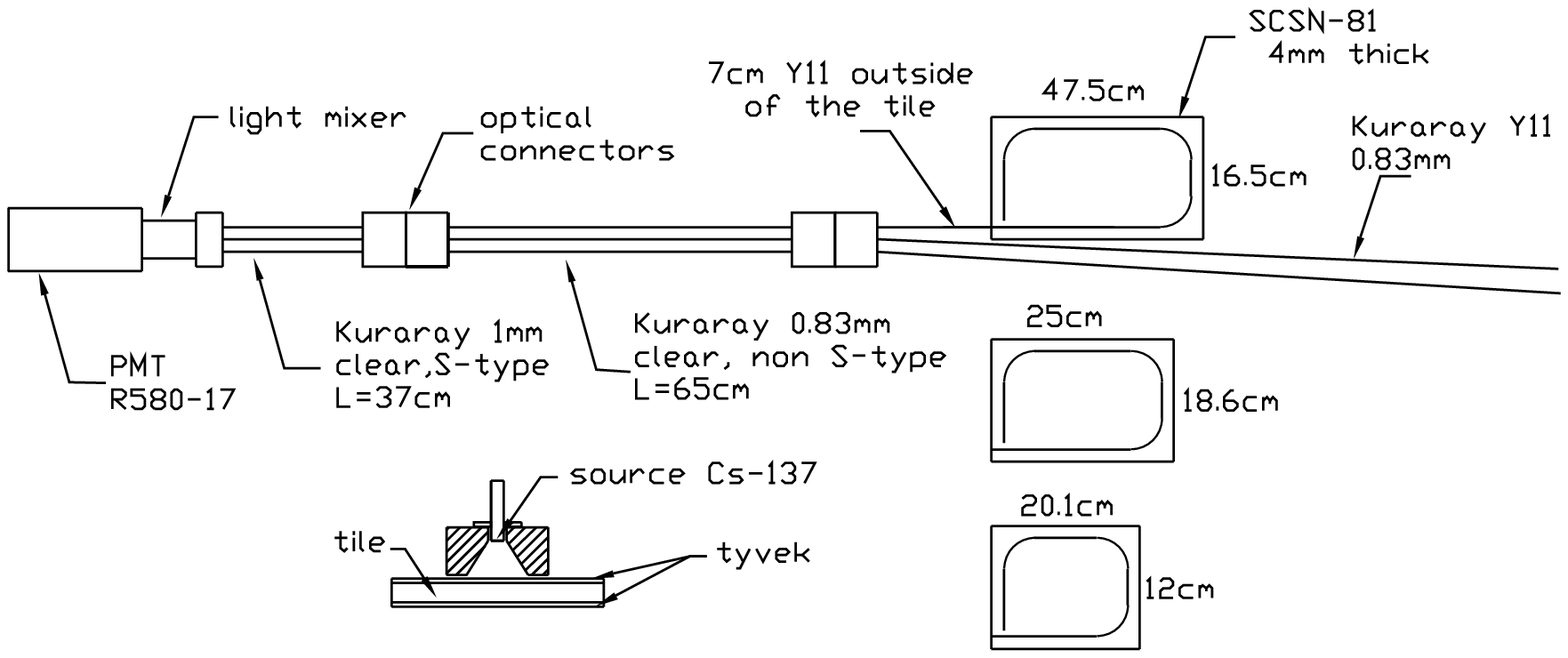}}\\
\label{light_vs_tile_size_lay_16_eva} 
\end{center}
\end{figure}

\begin{figure}
\begin{center}
\caption{Relative light vs tile size. (b) has the attenuation from the WLS
fiber between the tile and splice removed.}
\epsfxsize=2.7in
\mbox{{\epsffile[12 10 510 570]{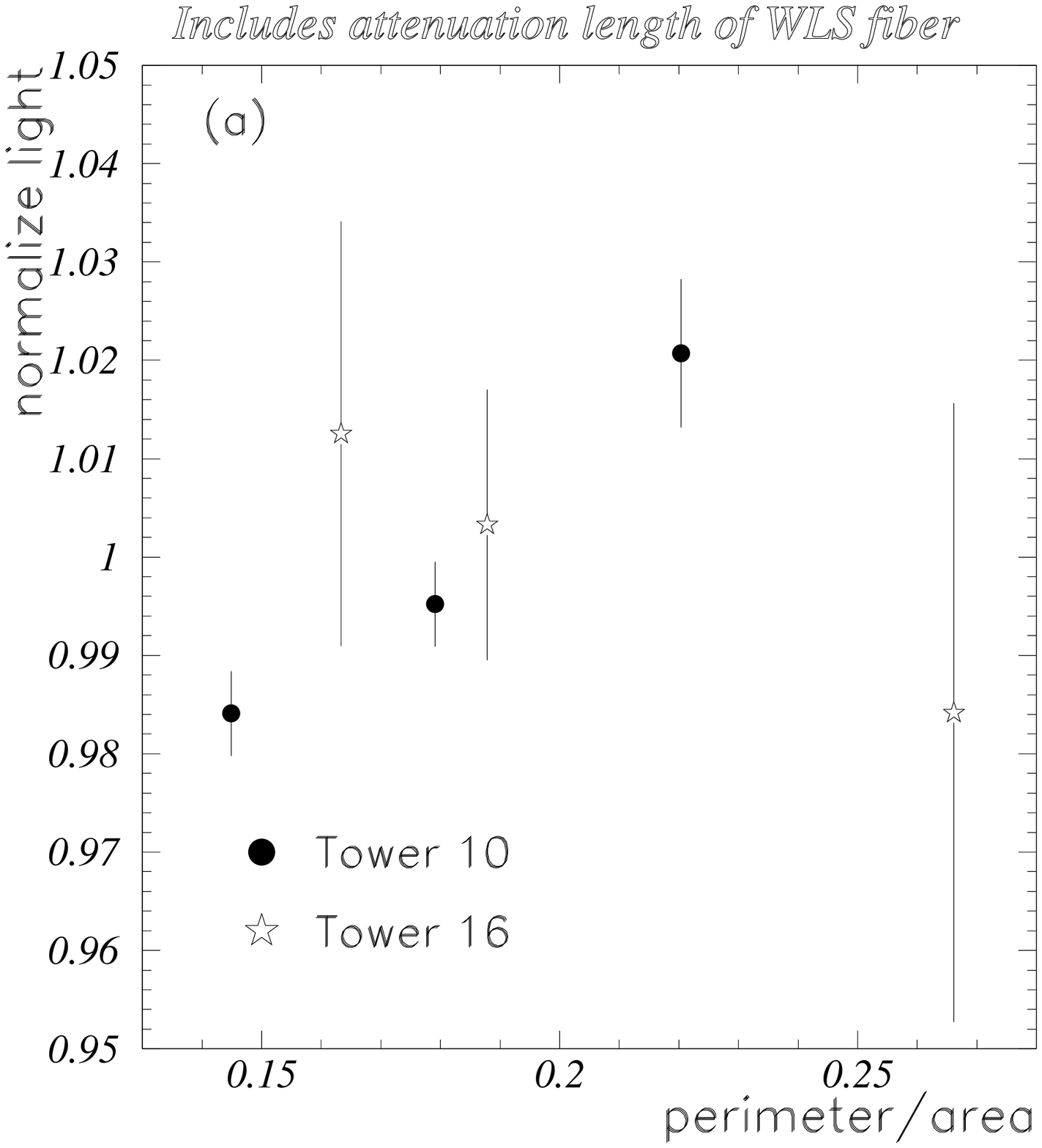}}
{\epsffile[12 10 510 570]{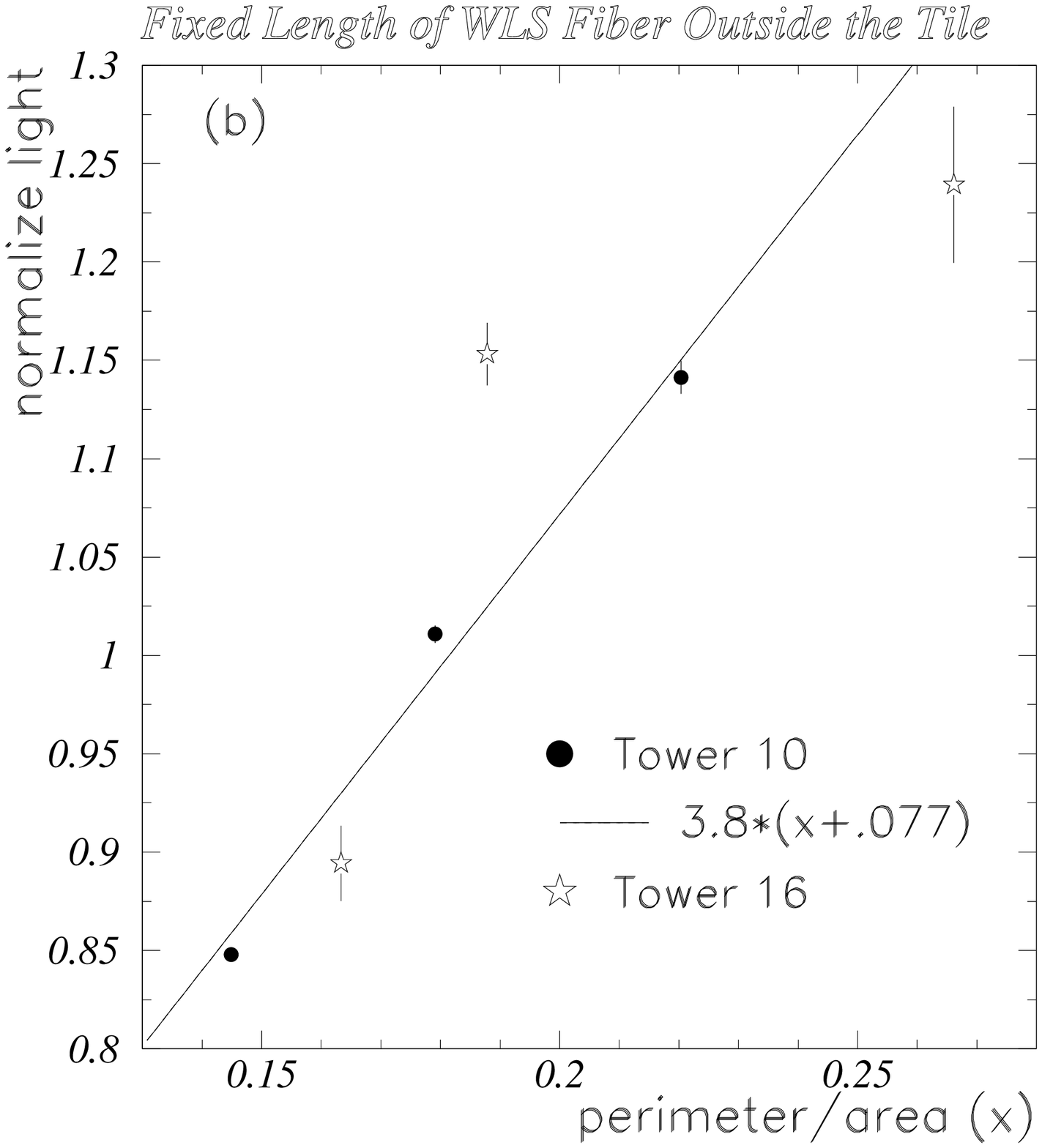}}}\\
\label{l_over_a}
\end{center}
\end{figure}

    Figure \ref{light_vs_tile_size_lay_10_eva}  and Figure
\ref{light_vs_tile_size_lay_16_eva} show the setup used to
measure the light vs tile size for CMS HCAL 
tower 10 and 16, respectively.  Tower 
10 is a tower in the middle of the barrel, while tower 16 is at the high
eta edge of the barrel. To measure tower 10,  we make 6 tiles each
for layers 1, 7, and 16, for a total of 18 tiles. 
We use the same pigtail to  measure  the
light from each of the tiles. The results are plotted as light vs 
perimeter/area \cite{lovera1} \cite{lovera2}. The
perimeter  measures the length of fiber in the tile, since the
distance the fiber  groove is from the edge of the scintillator is
kept at 0.3 cm  for these measurements. Figure \ref{l_over_a}a gives
the result. This result includes the additional length of WLS fiber
outside the tile for layers 1 and 7.  Figure \ref{l_over_a}b gives
the result with the WLS fiber attenuation removed. Figure
\ref{l_over_a}b gives  the light vs tile size. The overall
normalization is not measured by this measurement. The mean of the
data for tower 10 is normalized to 1  for both Figure
\ref{l_over_a}a and Figure \ref{l_over_a}b. The data on Figure
\ref{l_over_a}b is fit to a straight line.  The 3.8 
is a relative number that depends on the how the
light measurement is normalized.  The measurement measures the coefficent 
0.077/cm, which is the intercept divided by the  slope of the line. 

    To measure tower 16, we make 5 tiles each for layers 1, 5, and
8. Again,  we use the same pigtail to  measure the light from each
of the tiles. The pigtails for the tower 10 measurement and tower 16
measurement are different. Hence, the normalization of the tower 10
and tower  16 measurements are independent. The mean of the data 
for tower 16 in Figure \ref{l_over_a}a is set to 1. For Figure
\ref{l_over_a}b  the normalization for tower 16 
is set so that the mean of the
perimeter/area and mean of the normalized light lie on the straight
line for tower 10.  This enables us to see how consistent the two
measurements are.
 
  For the CMS design we used two models for the variation of light vs 
tile size. The first model assumes the light yield is a 
linear function of perimeter/area. The line is given in 
Figure \ref{l_over_a}b. We notice that the points for Tower 16 do
not follow a straight line. In the measurement for Figure \ref{l_over_a}a 
the same fibers are used to measure all the tiles in a tower.
 The total variation of the points for
Figure \ref{l_over_a}a for both Tower 10 and 16 is 4\%. The second
model assumes that a fiber with the same length green inserted 
into all the CMS tiles gives the same light. It assumes
the change in the light by changing the tile size is compensated for
by the attenuation in the green fiber.

\begin{figure}
\begin{center}
\caption{Setup to measure absolute light from a tile in CMS.
Results are given in Figure \protect\ref{abs_light}.}
\epsfxsize=5.7in
\mbox{\epsffile[20 40 560 194]{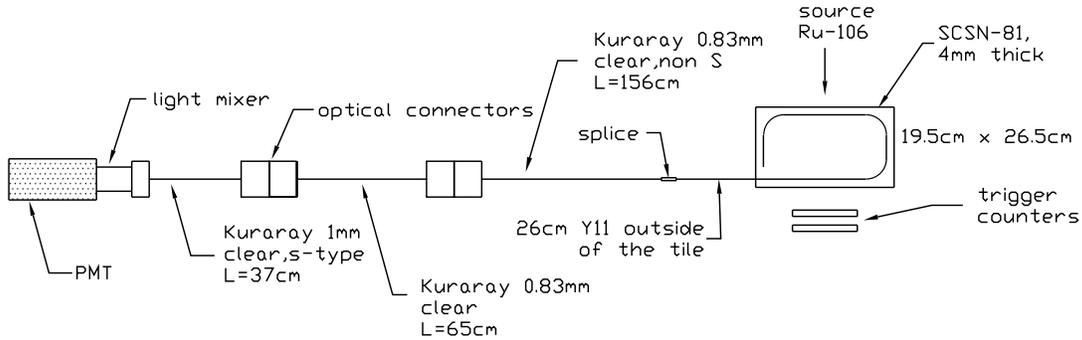}}\\
\label{abs_lite_eva} 
\end{center}
\end{figure}

\begin{figure}
\begin{center}
\caption{Absolute light yield.}
\epsfxsize=2.7in
\mbox{{\epsffile[12 20 510 550]{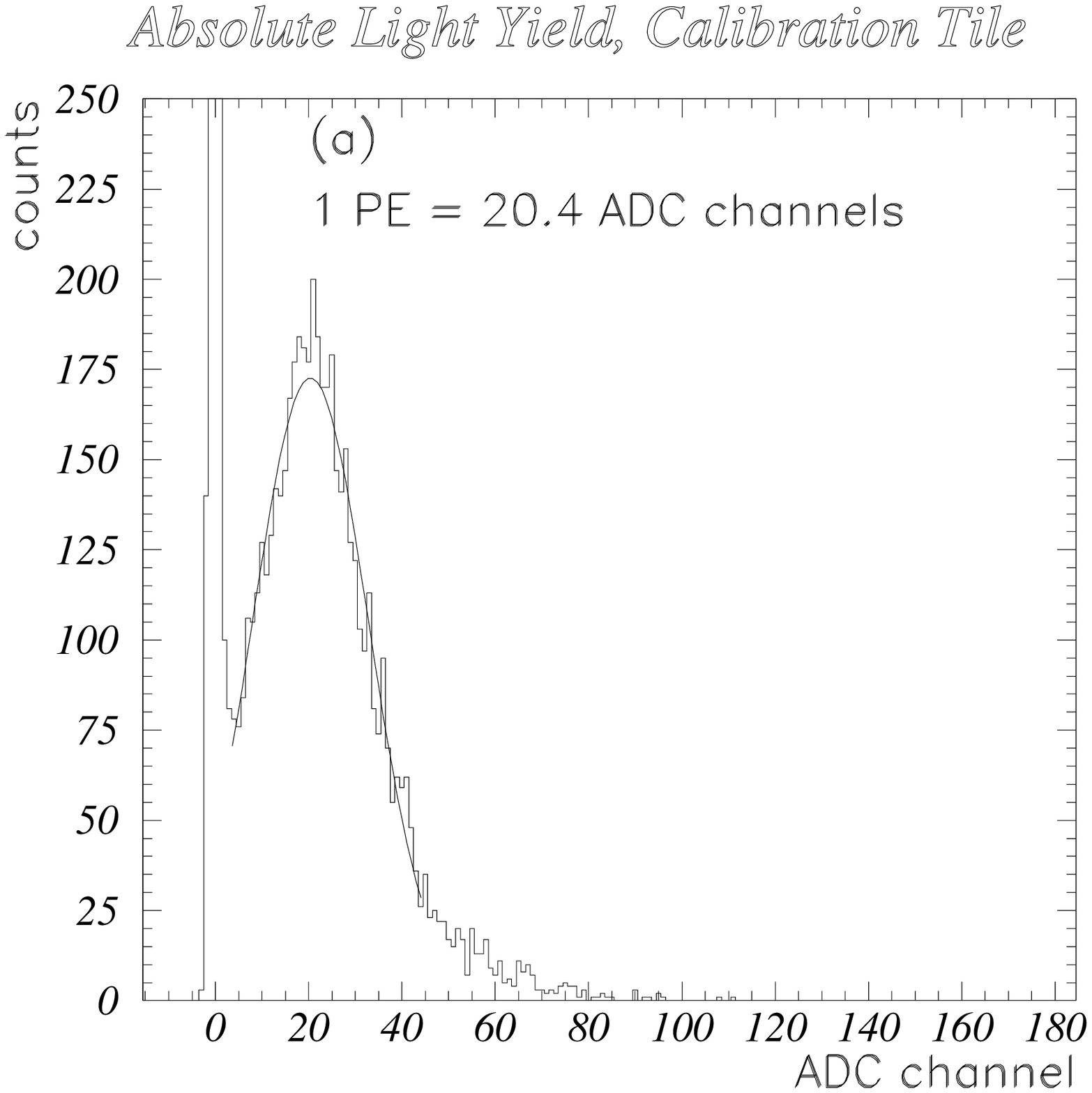}}
{\epsffile[12 20 510 550]{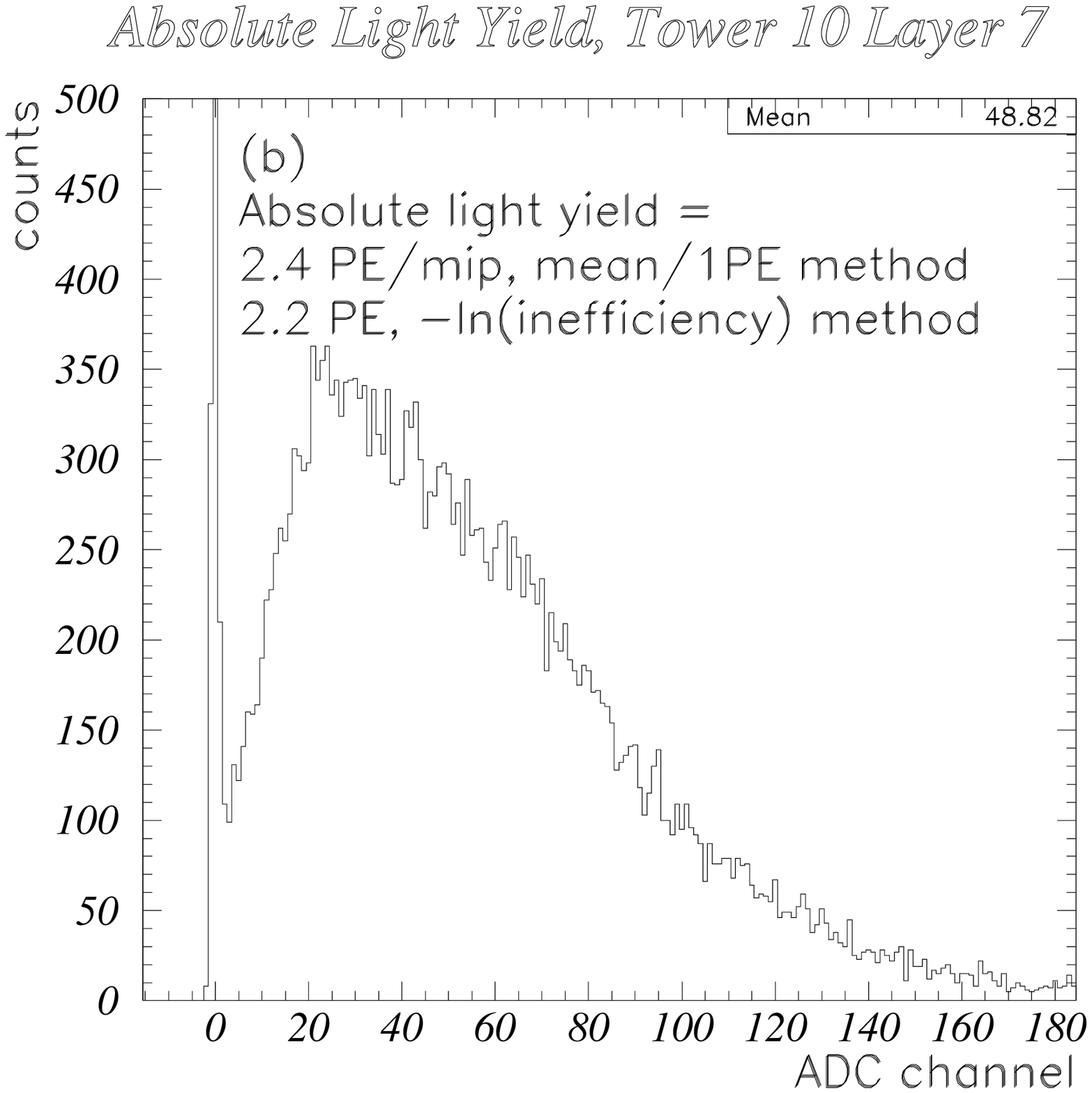}}}\\
\label{abs_light}
\end{center}
\end{figure}

    Figure 22 \ref{abs_lite_eva} shows the apparatus used to measure 
the absolute light yield of CMS tiles. A Ru-106 $\beta$ 
source is used instead of the Cs-137 source.  
Figure~\ref{abs_light} gives the result. 
A CMS HCAL barrel tile gives roughly 2 photoelectrons 
at the photodetector with a green extended photocathode.

The light for the tiles in HCAL barrel can be predicted 
using the attenuation length of the green and clear fiber, the 
model of the light vs tower size, and the absolute light yield. 
From this we can get the total light of a tower. All layers
of a tower, except for the first, go to the same photodetector. 
The longitudinal variation of light within a tower should be less than 10\%. 
By varying the position of the splice, we make the light 
uniform longitudinally in a tower.

\section*{Transverse Tile Uniformity}

\begin{figure}
\begin{center}
\caption{Apparatus to measure transverse tile uniformity with
1 mm air gap between the tiles. The fibers are inserted parallel 
to the short side. The result is shown in 
Figure \protect\ref{tile_plot_cms_short}.} 
\epsfxsize=5.7in
\mbox{\epsffile[30 150 550 360]{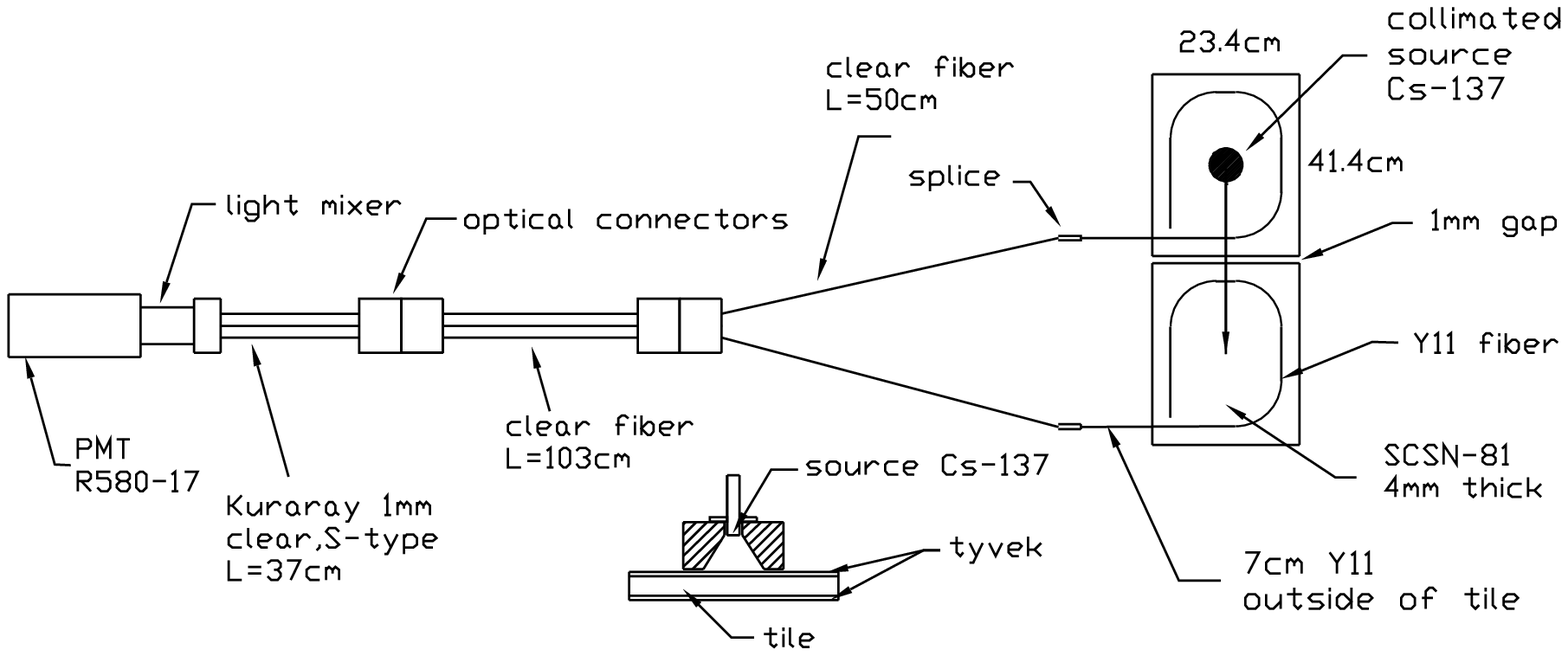}}\\
\label{tile_unif_cms_short_eva}
\end{center}
\end{figure}

\begin{figure}
\begin{center} 
\caption{Apparatus to measure transverse tile uniformity with 
1 mm air gap between the tiles. The fibers are inserted
into the tile parallel to the scan.
The measurement is shown in Figure \protect\ref{tile_plot_cms_long}.}
\epsfxsize=5.7in
\mbox{\epsffile[35 110 550 286]{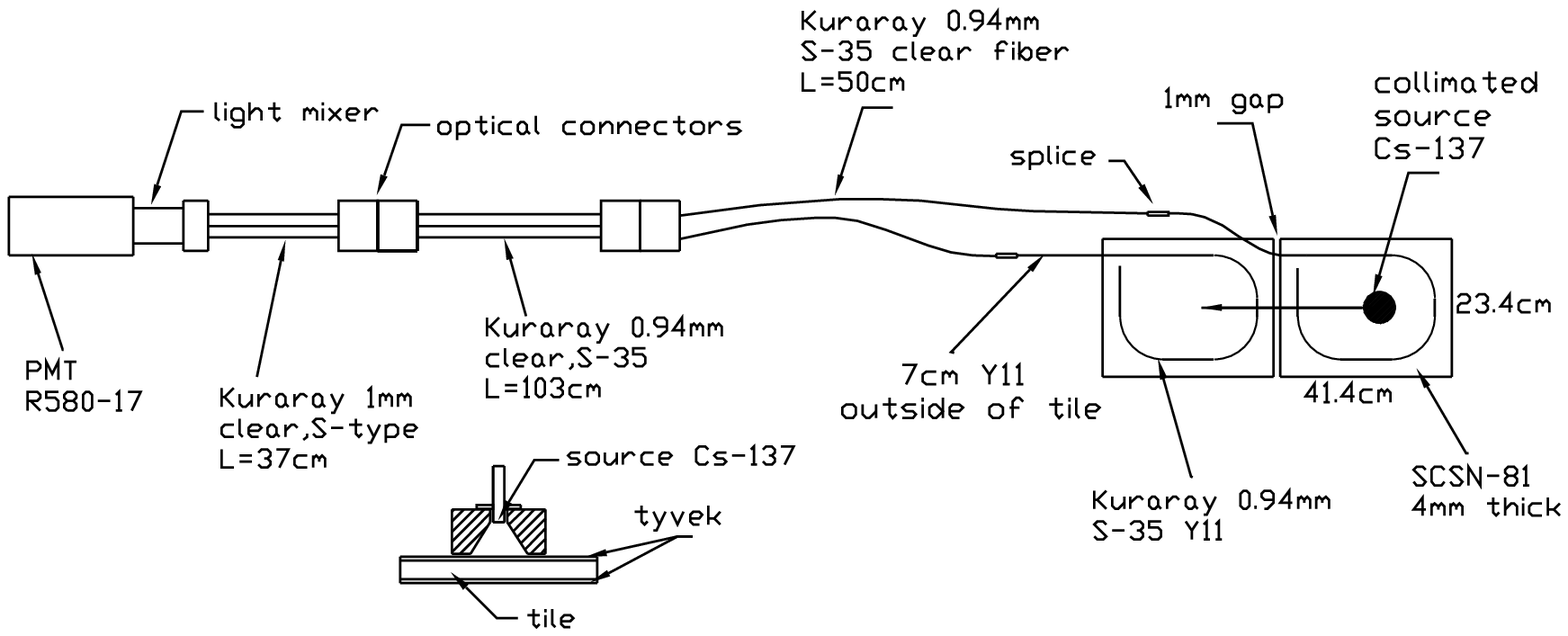}}\\
\label{tile_unif_cms_long_eva}
\end{center}
\end{figure}

  We have studied the transverse uniformity of the tiles. 
We constructed four tiles with the dimensions of tower 14, layer 14, 
which is the largest tile in the CMS HCAL Barrel. 
Two tiles have the fibers inserted parallel to the short side, 
called short side fiber insertion. 
The tiles with short side fiber
insertion are shown in Figure \ref{tile_unif_cms_short_eva}.
Two tiles have the fibers inserted parallel to the long side,
called long side fiber insertion. The tiles for long side fiber
insertion are shown in Figure \ref{tile_unif_cms_long_eva}.
CMS HCAL tiles have long side fiber insertion. The edges of the 
tiles are painted with white TiO$_2$ paint \cite{white_paint}.

\begin{figure}
\begin{center}
\caption{ Uniformity across a tile using collimated $\gamma$ source.
The fiber insertion is parallel to the short side. The measurement is
done with both Kuraray and Bicron fibers.}
\epsfxsize=2.7in
\mbox{{\epsffile[12 10 510 550]{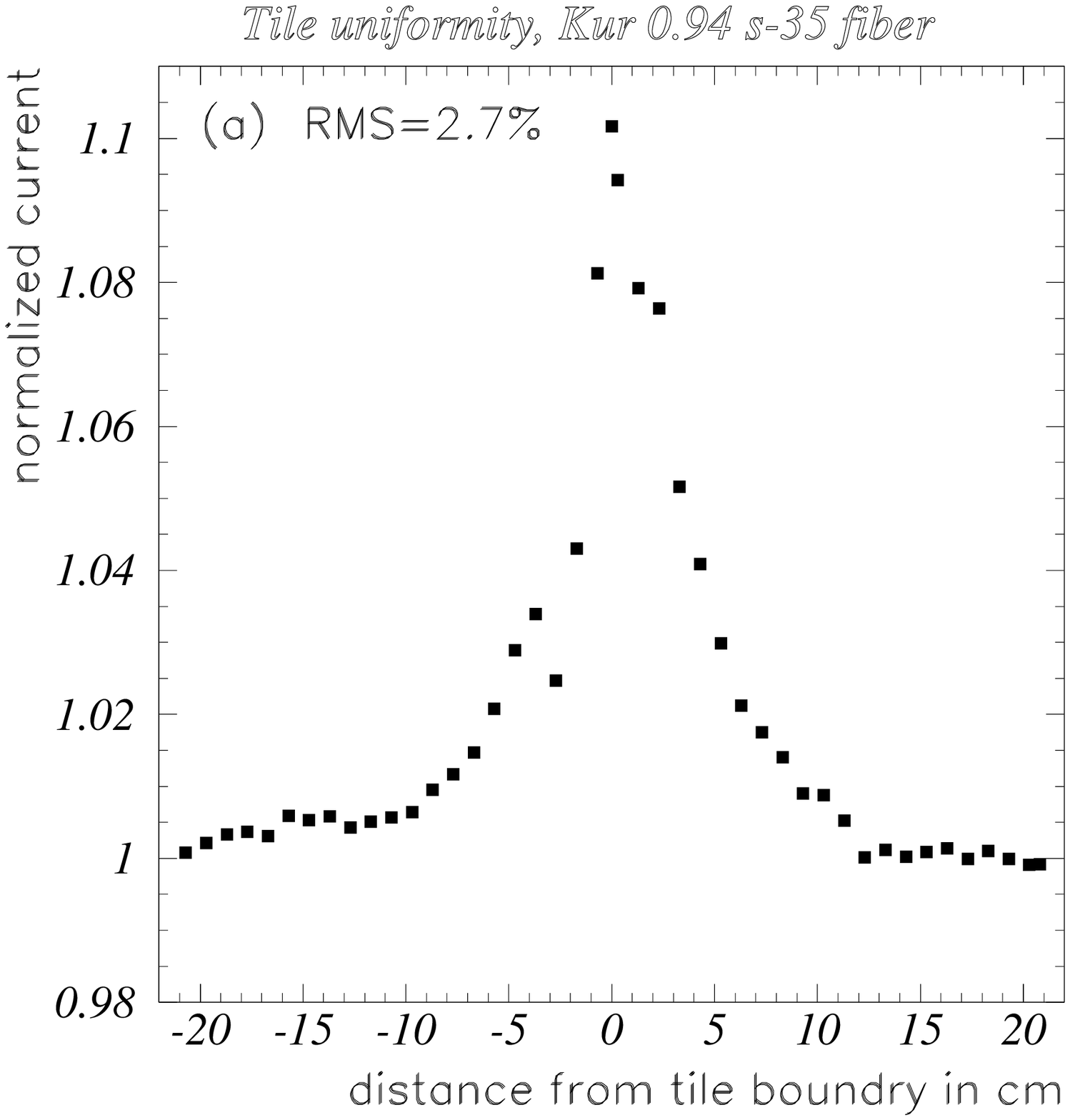}} 
{\epsffile[12 30 510 550]{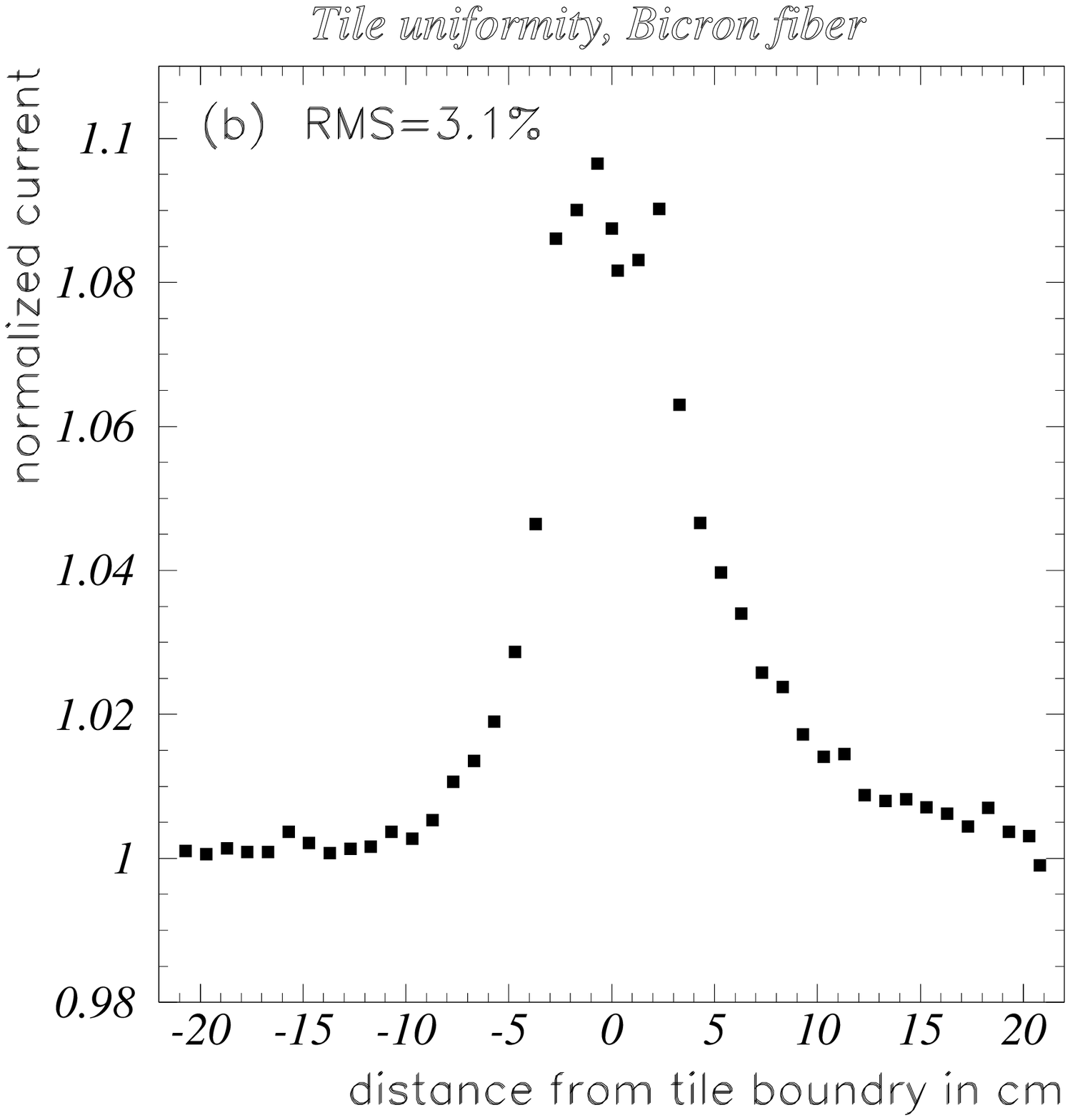}}}\\
\label{tile_plot_cms_short}
\end{center}
\end{figure}

\begin{figure}
\begin{center}
\caption{Uniformity across a tile using collimated $\gamma$ source.
The fiber insertion is parallel to the long side.}
\epsfxsize=2.7in
\mbox{\epsffile[12 20 510 550]{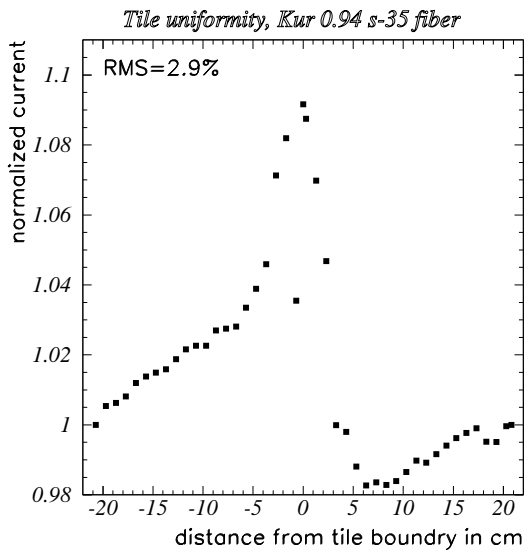}} 
\label{tile_plot_cms_long}
\end{center}
\end{figure}

The uniformity is measured with a collimated Cs-137 $\gamma$ source. 
The collimator constrains the radiation to a 7.5 cm diameter circle. 
The central transverse size of a hadron shower is approximately 7.5 cm. 
Hence, the collimated source  
simulates the transverse size of a hadron shower.
Figure \ref{tile_plot_cms_short} shows the uniformity across the 
tile with the short side fiber insertion. 
The uniformity was measured with both Kuraray fibers and Bicron 
fibers. The Kuraray measurement uses 0.94 mm S-35 fibers for both
the pigtail and the cable. For the Bicron measurement, the WLS  
1.0 mm Bicron fiber (Batch 2) is spliced to 1.0 mm
non-S Kuraray fiber. The cable for the Bicron measurement was
made with S type 1.0 mm Kuraray fiber. 
For both kinds of fibers the tile is very uniform 
with a 10\% increase at the boundary between the 2 tiles. 
The increase is due to increased light collection
at the fiber.   Figure \ref{tile_plot_cms_long} shows 
the uniformity  with long side fiber insertion.
The transverse uniformity has an RMS $\sim$ 3\% regardless
of the fiber type or fiber insertion point.
The resolution of the CMS calorimeter is $120\%/\sqrt{E}~\oplus~5\%$ 
\cite{cms_calorimeter_talk}.  The transverse uniformity across the
tile should be somewhat less than the constant term, 
5\%, to prevent transverse uniformity
from affecting the constant term in the resolution.

\begin{figure}
\begin{center}
\caption{The boundary between tiles in CMS HCAL barrel.} 
\epsfxsize=5.7in
\mbox{\epsffile[50 130 550 360]{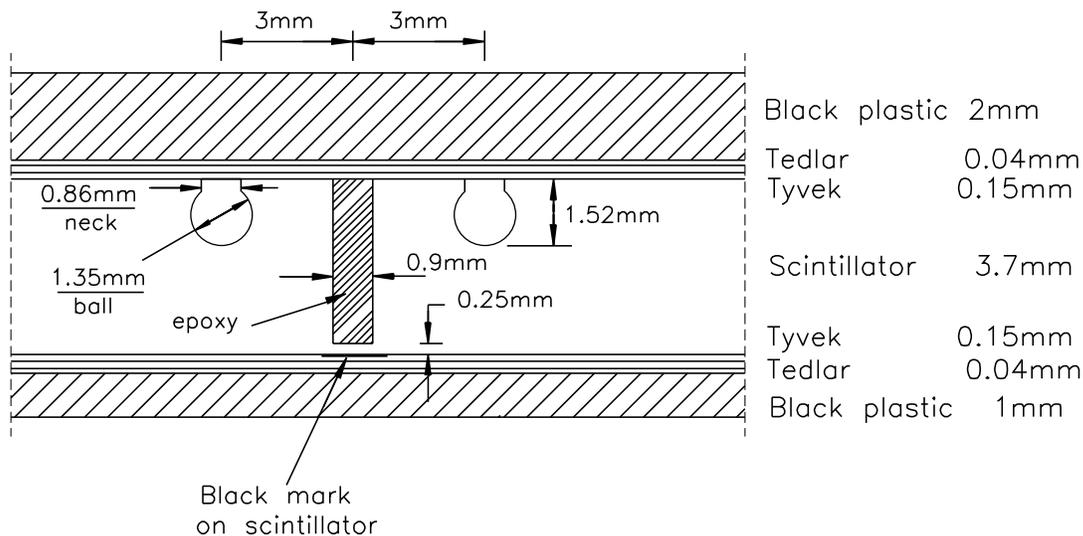}}\\
\label{cms_tile_boundary_eva}
\end{center}
\end{figure}

 The HCAL CMS scintillator design has individual tiles 
glued together with TiO$_2$ loaded epoxy resin
\cite{glue}, to form a "megatile". 
The configuration at the boundary between tiles is 
shown in Figure \ref{cms_tile_boundary_eva}. 
A 0.9 mm wide "separation groove" 
is cut to separate 2 tiles, with 1/4 mm of 
scintillator left uncut on the bottom of the groove. The groove is filled 
with TiO$_2$ loaded epoxy.  
The scintillator is marked with a black mark made
with the narrow end of a black marker pen 
\cite{doubleshot}. The black mark is underneath the separation groove.
The black mark is about 1.5 mm wide, slightly wider than the separation groove. 
The black mark reduces the light cross talk 
through the 1/4 mm of scintillator left at the tile boundary.

\begin{figure}
\begin{center}
\caption{Apparatus to measure transverse tile uniformity with
1 mm glue joint between the tiles.} 
\epsfxsize=5.7in
\mbox{\epsffile[40 60 550  223]{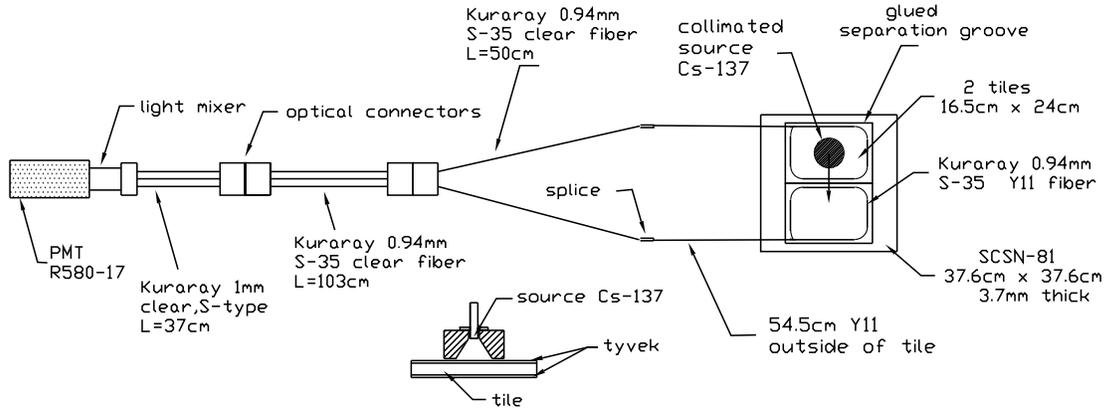}}\\
\label{tile_unif_glue_tile_eva}
\end{center}
\end{figure}

\begin{figure}
\begin{center}
\caption{ Uniformity across the glued megatile 
using collimated $\gamma$ source.}
\epsfxsize=2.7in
\mbox{\epsffile[12 40 510 550]{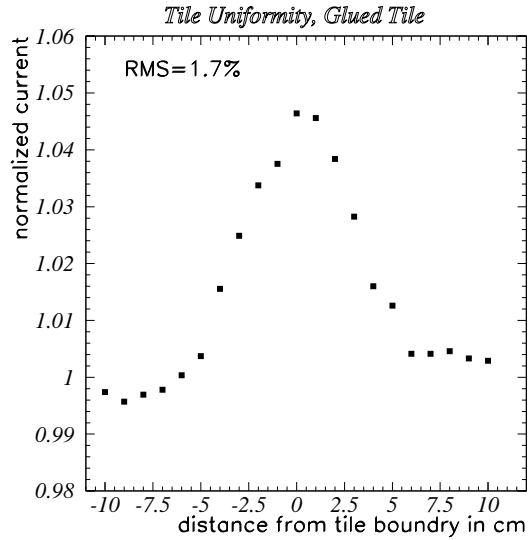}} 
\label{tile_plot_mark}
\end{center}
\end{figure}

We constructed a glued megatile consisting of 2 tiles inside
a piece of scintillator.
The tiles for that glued megatile tile are from tower 10, layer 1. 
Figure \ref{tile_unif_glue_tile_eva} shows the apparatus 
used to measure the glued megatile. Figure \ref{tile_plot_mark}
shows the result. 
The transverse RMS is roughly 1.7\%. The 
transverse uniformity does not increase the 
constant term of the resolution. 

We have measured the cross talk between the glued tiles.
The 1/4 mm of scintillator, left uncut between the two tiles,
provides a path for light to pass between 2 tiles.
The cross talk is measured by first putting a fiber in one of the tiles.
Next, we measure the current with the source on the following
three locations: the tile with the fiber, the tile without 
the fiber, and just off the tile with the fiber. The last 
location is used  to measure the source 
cross talk.  The cross talk is $\sim$ 1\%. 

The black mark decreases the light output.
We measured the light from glued megatile. Next, the
black marks are made on 3 sides of the tile, similar
to the way the HCAL CMS barrel tiles will be marked.
The light goes down roughly 8\%.

\section*{Conclusion}

The CMS R and D enables us to design the optics of HCAL
Barrel Calorimeter and predict its performance. We have
chosen Kuraray S-35 fiber for the HCAL preproduction prototype because of
its excellent flexibility, excellent mirror reflectivity, and
high splice transmission. CMS HCAL has chosen to ice polish 
the fibers, since it enables us
to polish many fibers at once. We predict the light of each
tile in the barrel using the attenuation lengths of fibers
and the absolute light vs the tile size. By varying the 
position of the splice for each tile, we can optimize the 
light distribution in a tower.  CMS has chosen to have the same
length WLS fiber for all layers in a tower. 
Measurements of the transverse uniformity shows that it 
does not effect the resolution of the calorimeter.
%

\end{document}